\documentclass[a4paper,fleqn]{cas-sc}

\usepackage[numbers]{natbib}
\usepackage[separate-uncertainty, multi-part-units=single]{siunitx} % for unit
\def\tsc#1{\csdef{#1}{\textsc{\lowercase{#1}}\xspace}}
\tsc{WGM}
\tsc{QE}
\begin{document}
\let\WriteBookmarks\relax
\def\floatpagepagefraction{1}
\def\textpagefraction{.001}

%Short title
\shorttitle{The INTT Detector for sPHENIX}    

\shortauthors{<Y. Akiba, {\it et al.}>}  

\title [mode=title]{The Intermediate Silicon Tracker Detector for sPHENIX}  


%\tnotetext[1]{This work is supported by } 


\author[1,2]{Y.~Akiba}[] % Yasuyuki Akiba
\author[7,1]{H.~Aso}[] % Hiromu Aso
\author[9,2]{J.~T.~Bertaux}[] % Joseph T. Bertaux
\author[3]{D.~Cacace}[] % Daniel Cacace
\author[3]{R.~G.~Cecato} % Raul G. Cecato
\author[4]{K.~Y.~Chen}[] % Ke-Yi Chen
\author[4,2]{S.~Y.~Chen}[] % Shan-Yu Chen
\author[4,2]{K.~Y.~Cheng}[] % Kai-Yu Cheng
\author[1]{A.~Enokizono}[]\cormark[1]\ead{enoki@ribf.riken.jp} % Akitomo Enokizono
\author[1,2]{H.~Enyo}[] % Hideto Enyo
\author[7,1]{K.~Fujiki}[] % Kazuma Fujiki
\author[7,1]{Y.~Fujino}[] % Yusuke Fujino
\author[5,1]{M.~Fujiiwara}[] % Manami Fujiwara
\author[5,2]{T.~Hachiya}[] % Takashi Hachiya
\author[7,1]{T.~Harada}[] % Tomoki Harada
\author[6]{S.~Hasegawa}[] % Shoichi Hasegawa
\author[5,2]{M.~Hata}[] % Misaki Hata
\author[11]{B.~Hong}[] % Byungsik Hong
\author[11,2]{J.~Hwang}[] % Jaein Hwang
\author[7,1]{T.~Ichino}[] % Takahiro Ichino
\author[5,1]{M.~Ikemoto}[] % Mahiro Ikemoto
\author[7,1]{D.~Imagawa}[] % Daisuke Imagawa
\author[7,1,2]{H.~Imai}[] % Hikaru Imai
\author[5,1]{Y.~Ishigaki}[] % Yui Ishigaki
\author[5]{M.~Isshiki}[]  % Moe Isshiki
\author[5,1]{K.~Iwatsuki}[] % Koyuki Iwatsuki
\author[5]{R. Kan}[] % Rin Kan
\author[5,1]{M.~Kano}[] % Mai Kano
\author[7,1]{T.~Kato}[] % Tomoya Kato
\author[7,1]{R.~Kawashima}[] % Ryota Kawashima
\author[7,1]{T.~Kikuchi}[] % Takahiro Kikuchi
\author[12,1]{R.~Koike}[] % Ryotaro Koike
\author[8]{T.~Kondo}[] % Takashi Kondo
\author[4]{C.~M.~Kuo}[] % Chia-Ming Kuo
\author[5]{H.~Kureha}[] % Hiroko Kureha
\author[1]{T.~Kumaoka}[] % Takuya Kumaoka
\author[9]{H.~S.~Li}[] % Han-Sheng Li 
\author[10]{R.~S.~Lu}[] % Rong-Shyang Lu
\author[3]{E.~Mannel}[] % Eric Mannel
\author[7,1]{H.~Masuda}[] % Hidekazu Masuda
\author[2]{G.~Mitsuka}[] % Gaku Mitsuka
\author[5,1]{N.~Morimoto}[] % Nao Morimoto
\author[5,2]{M.~Morita}[] % Miu Morita
\author[1,2]{I. Nakagawa}[] % Itaru Nakagawa
\author[7,1]{Y.~Nakamura}[] % Yusuke Nakamura
\author[7,1]{G.~Nakano}[] % Genta Nakano
\author[5,2]{Y.~Namimoto}[] % Yumika Namimoto
\author[12]{M.~Naruki}[] % Megumi Naruki
\author[7,1]{D.~Nemoto}[] % Daisuke Nemoto
\author[5]{S.~Nishimori}[] % Sakiko Nishimori
\author[3]{R.~Nouicer}[] % Rachid Nouicer
\author[6,1]{G.~Nukazuka}[] % Genki Nukazuka
\author[5,1]{I.~Omae}[] % Itsuka Omae
\author[3]{R.~Pisani}[] % Robert Pisani
\author[1]{Y.~Sekiguchi}[] % Yuko Sekiguchi
\author[5,2]{M.~Shibata}[] % Mika Shibata
\author[4,2]{C.~W.~Shih}[] % Cheng-Wei Shih
\author[7,1]{K.~Shiina}[] % Kenichiro Shiina
\author[5]{M.~Shimomura}[] % Maya Shimomura
\author[7,1]{R.~Shishikura}[] % Ryota Shishikura
\author[9,2]{M.~Stojanovic}[] % Milan Stojanovic
\author[5]{K.~Sugino}[] % Kazune Sugino
\author[5]{Y.~Sugiyama}[] % Yuka Sugiyama
\author[5,2]{A.~Suzuki}[] % Ayaka Suzuki
\author[5,2]{R.~Takahama}[] % Runa Takahama
\author[4,2]{W.~C.~Tang}[] % Wei-Che Tang
\author[5]{Y.~Terasaka}[] % Yuri Terasaka
\author[2]{T.~Todoroki}[] % Takahito Todoroki
\author[13]{N.~Tomida}[] % Natsuki Tomida
\author[5,1]{H.~Tsujibata}[] % Hinako Tsujibata
\author[7,1]{M.~Tsuruta}[] % Masato Tsuruta
\author[2]{Y.~Yamaguchi}[] % Yorito Yamaguchi
\author[7,1]{H.~Yanagawa}[] % Hayato Yanagawa
\author[5,2]{M.~Watanabe}[] % Mai Watanabe
\author[9]{R.~Xiao}[] % Rui Xiao
\author[9]{W.~Xie}[] % Wei Xie
%Total 47 institutes




%\ead[url]{http://riken.jp}

%\credit{<Credit authorship details>}

\affiliation[1]{organization={Nishina Center for Accelerator-Based Science, RIKEN},
            addressline={2-1 Hirosawa}, 
            city={Wako},
            postcode={351-0198}, 
            state={Saitama},
            country={Japan}}

\affiliation[2]{organization={RIKEN BNL Research Center},
            addressline={20 Pennsylvania Avenue}, 
            city={Upton},
            postcode={11973}, 
            state={NY},
            country={U.S.A.}}            
            
\affiliation[3]{organization={Brookhaven National Laboratory},
            addressline={20 Pennsylvania Avenue}, 
            city={Upton},
            postcode={11973}, 
            state={NY},
            country={U.S.A.}}

\affiliation[4]{organization={Department of Physics and Center for High Energy and High Field Physics, National Central University},
            addressline={No.300, Zhongda Rd., Zhongli Dist.}, 
            city={Taoyuan City},
            postcode={32001}, 
            country={Taiwan}}

\affiliation[5]{organization={Department of Mathematical and Physical Sciences, Nara Women's University},
            addressline={Kitauoya-Higashimachi}, 
            city={Nara},
            postcode={630-8506}, 
            state={Nara},
            country={Japan}}         
            
\affiliation[6]{organization={Advanced Science Research Center, Japan Atomic Energy Agency},
            addressline={2-4 Shirakata Shirane}, 
            city={Tokai-mura, Naka-gun},
            postcode={319-1195}, 
            state={Ibaraki},
            country={Japan}}
            
\affiliation[7]{organization={Department of Physics, Rikkyo University},
             addressline={3-34-1 Nishi-Ikebukuro}, 
            city={Toshima-ku},
            postcode={171-8501}, 
            state={Tokyo},
            country={Japan}} 
            
\affiliation[8]{organization={Tokyo Metropolitan Industrial Technology Research Institute},
             addressline={2-4-10, Aomi}, 
            city={Koto-ku},
            postcode={135-0064}, 
            state={Tokyo},
            country={Japan}} 

\affiliation[9]{organization={Department of Physics and Astronomy, Purdue University},
             addressline={525 Northwestern Ave.}, 
            city={West Lafayette},
            postcode={47907}, 
            state={IN},
            country={U.S.A.}} 
            
\affiliation[10]{organization={Department of Physics, National Taiwan University},
             addressline={No.1 Sec.4 Roosevelt Road}, 
            city={Taipei},
            postcode={10617}, 
            country={Taiwan}} 
            
\affiliation[11]{organization={Department of Physics, Korea University},
             addressline={Anam-dong 5, Seongbuk-gu}, 
            city={Seoul},
            postcode={02841}, 
            country={Korea}} 

\affiliation[12]{organization={Department of Physics, Kyoto University},
             addressline={Kitashirakawa Oiwake-cho, Sakyo-ku}, 
            city={Kyoto},
            postcode={606-8502}, 
            state={Kyoto},
            country={Japan}} 

\affiliation[13]{organization={Center for Science Adventure and Collaborative Research Advancement, Kyoto University},
             addressline={Kitashirakawa Oiwake-cho, Sakyo-ku}, 
            city={Kyoto},
            postcode={606-8502}, 
            state={Kyoto},
            country={Japan}}

% Footnote of the second author

% Email id of the second author
%\ead{}



\cortext[1]{Corresponding author at: Nishina Center for Accelerator-Based Science, RIKEN, 2-1 Hirosawa, Wako, 351-0198, Saitama, Japan.}

%\fntext[1]{xxxx}


\begin{abstract}
A new barrel-type silicon detector has been constructed in the central rapidity region of the sPHENIX experiment as part of its charged-particle tracking system.
The detector, referred to as the intermediate silicon tracker (INTT), provides single-event timing capability and contributes to charged-particle tracking in combination with the silicon-pixel vertex detector and the time-projection chamber.
The INTT consists of two cylindrical barrels with full azimuthal coverage. The inner and outer barrels comprise 24 and 32 silicon strip ladders, respectively. Each ladder consists of four silicon strip sensors, covering the pseudorapidity range $|\eta| < 1.07$ for collision vertices within $\pm~10$ cm along the beam axis, at radial distances between 72 and 103 mm.
The INTT has been fully integrated into the sPHENIX data acquisition system, with detector signals processed by eight FELIX servers. Continuous data taking is enabled by a streaming readout mode, in which detector data are accumulated without hardware triggers, while a conventional trigger-based mode is also supported. The detector achieves stable operation, small inactive area, and high efficiency, providing precise timing, tracking, and vertexing performance during the collection of beam-beam collision data.

\end{abstract}

%\item INTT
%\item sPHENIX
%\item RHIC

% Keywords

\maketitle

% Main text

%%%%%%%%%%%%%%%%%%%%%%%%%%%%%%%%%%%%%%%%%%%%%%%%%%
%%%%%%%%%%%%%%%% Detecor Overview %%%%%%%%%%%%%%%%
%%%%%%%%%%%%%%%%%%%%%%%%%%%%%%%%%%%%%%%%%%%%%%%%%%
\section{Introduction}
\label{Introduction}
The sPHENIX detector~\cite{Aidala:2012nz}, located at the Relativistic Heavy Ion Collider (RHIC)~\cite{Hahn:2003nim} at the Brookhaven National Laboratory (BNL), is a major upgrade of the PHENIX detector~\cite{Adcox:2003nim}, which ceased operation in 2016. The sPHENIX experiment is designed to record high-statistics proton–proton (p+p) and nucleus–nucleus collisions, enabling precision studies of jet modification, quarkonium suppression, and open heavy-flavor production. These measurements probe the properties of the strongly coupled quark–gluon plasma~\cite{Belmont:2023fau} and are complementary to those performed at the Large Hadron Collider~\cite{LRP_2015_cite}.

\begin{figure}[htb]
\begin{center}
\includegraphics[width=0.48\textwidth]{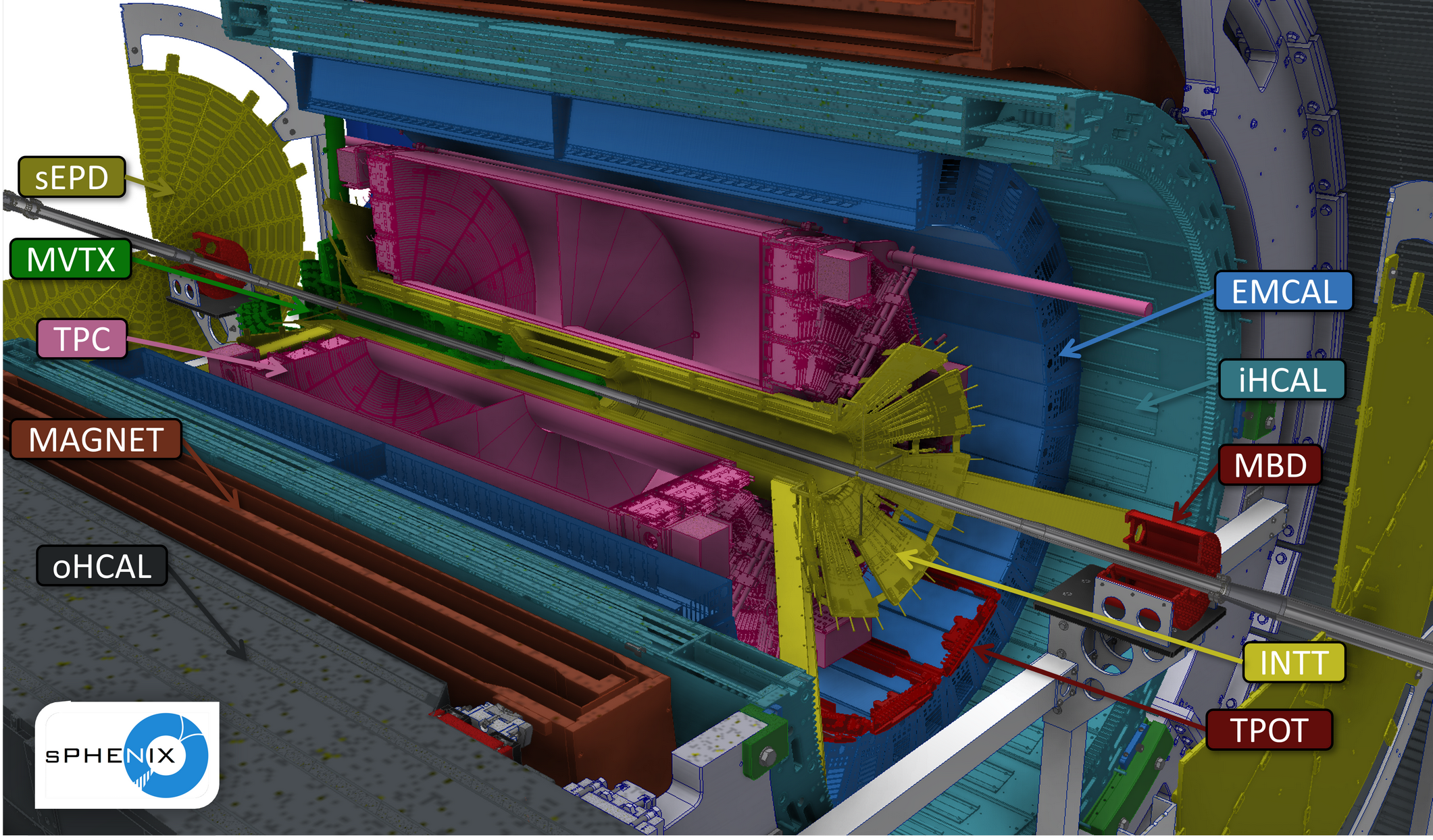}
\caption{
The mechanical drawing of the sPHENIX detector.
\label{fig:sPHENIX}
}
\end{center}
\end{figure}

The detector provides precision vertexing, tracking, and electromagnetic and hadronic calorimetry in the central pseudorapidity region ($|\eta|<1.1$) with full azimuthal coverage at the nominal RHIC collision rate. The overall detector layout follows the general design of modern collider experiments and is shown in Fig.~\ref{fig:sPHENIX}. The tracking system consists of a monolithic active pixel sensor (MAPS) vertex detector (MVTX), an intermediate silicon tracker (INTT)~\cite{Nakagawa2025, SHIH2026171312}, and a time-projection chamber (TPC). A micromegas-based detector (TPOT)~\cite{Aune:2024nim} is installed outside the TPC for calibration purposes.
The calorimeter system comprises a tungsten/scintillating-fiber electromagnetic calorimeter (EMCal), an inner aluminum/scintillator hadronic calorimeter (iHCal), and an outer steel/scintillator tile hadronic calorimeter (oHCal). The iHCal is located inside a 1.4~T superconducting solenoid, refurbished from the BaBar experiment~\cite{Aubert2002nim}, and the oHCal is located outside the solenoid.

The sPHENIX detector started commissioning data collection in 2023 with gold-gold (Au+Au) collisions at $\sqrt{s_{\rm NN}} = 200$ GeV.
This paper describes the INTT readout configuration, the barrel structure, and its performance evaluated with collision data and cosmic rays. The performance of the pre-production version of the silicon ladders in a beam test is reported in~\cite{SHIH2026171312}.

%%%%%%%%%%%%%%%%%%%%%%%%%%%%%%%%%%%%%%%%%%%%%%%%%%
%%%%%%%%%%%%%%%% Detecor Overview %%%%%%%%%%%%%%%%
%%%%%%%%%%%%%%%%%%%%%%%%%%%%%%%%%%%%%%%%%%%%%%%%%%
\section{INTT Detector Overview}
\label{Overview}

The INTT plays a key role in the sPHENIX tracking system by providing precise timing information with superior timing resolution. This capability enables the association of reconstructed charged-particle tracks with their corresponding beam crossings, allowing discrimination of tracks originating from collisions in different bunch crossings. The MVTX provides excellent spatial resolution for precise vertex determination, while the TPC offers high momentum resolution. The combination of these three subsystems enables high-precision charged-particle track reconstruction. 

The barrel-type INTT detector consists of inner and outer layers of silicon ladders. Adjacent ladders are azimuthally staggered to eliminate inactive regions in the acceptance. The inner and outer barrels comprise 24 and 32 ladders, respectively.

A brief overview of the INTT ladders and their readout cables is presented in this paper. Detailed descriptions of INTT silicon ladders and readout cables are provided elsewhere~\cite{Nakagawa2025}.

Each silicon ladder is composed of silicon strip sensors designed for precise charged-particle detection. The sensors have a strip pitch of \SI{78}{\micro\meter}, a thickness of \SI{320}{\micro\meter}, and lengths of 16 or 20 mm~\cite{SiliconHPK}.

The design of the INTT ladders and their readout is based on the FPHX readout chip~\cite{PHENIX_FPHX_cite, Kapustinsky:2010nim, FPHX_Manual} and the readout card (ROC) developed for the forward silicon vertex tracker (FVTX) detector~\cite{Aidala:2013vna} in PHENIX. While newly fabricated FPHX chips are used for the INTT, the ROC modules are reused. Consequently, the overall INTT design was optimized to be compatible with the existing readout architecture. The primary difference between the FVTX and INTT detectors lies in their mechanical configurations: the INTT adopts a two-layer barrel geometry, whereas the FVTX employs a four-layer disk geometry.

Each INTT ladder module consists of two silicon strip sensors and 26 FPHX readout chips mounted on a high-density interconnect (HDI) flexible printed circuit. Two such modules are aligned along the beam direction and mounted on a carbon-fiber composite (CFC) support structure (stave) to form an INTT ladder. 
Thermal management is achieved using high thermal conductivity materials, including carbon fibers and adhesive layers, to efficiently spread the heat generated by the FPHX chips. Heat is removed via a water-cooling system implemented through a carbon cooling tube integrated into the stave.

The readout of each ladder is performed from both longitudinal ends, with each silicon module readout independently. The HDI transmits data from half of a ladder to downstream electronics. The data are further transferred via a bus extender (BEX) cable~\cite{BEX} and a conversion cable (CC)~\cite{Nakagawa2025}. The BEX is a 1.11~m flexible printed cable employing liquid-crystal polymer as a dielectric material. The conversion cable consists of three components: (1) a $\mu$-coaxial harness, (2) power and ground lines, and (3) connector interface boards. The downstream end of the conversion cable is connected to the ROC, which aggregates data from up to seven half-ladders and transmits formatted data via optical fiber links to a front-end link exchange (FELIX) server~\cite{Felix, Anderson_2016}, where the data are integrated into the sPHENIX data acquisition system.

%%%%%%%%%%%%%%%%%%%%%%%%%%%%%%%%%%%%%%%%%%%%%%%%%%
%%%%%%%%%%%%%%%% Barrel Assembly %%%%%%%%%%%%%%%%%
%%%%%%%%%%%%%%%% Rachid, Dan     %%%%%%%%%%%%%%%%%
%%%%%%%%%%%%%%%%%%%%%%%%%%%%%%%%%%%%%%%%%%%%%%%%%%
\section{INTT Barrel Description and Assembly}

The INTT barrel consists of INTT ladder modules, readout cables, services, and supports.
The specifications of the INTT ladder module and readout cable are described in~\cite{Nakagawa2025, BEX}, and this section describes the materials, geometry, and installation process for the support structure and the assembled INTT barrel.

\subsection{Design Drivers}

The INTT is positioned between the MVTX and the TPC inside the inner diameter of the sPHENIX solenoid magnet.  The function of the INTT is to provide high-precision tracking together with the MVTX and the TPC in the high-multiplicity environment of heavy-ion collisions, while complying with strict mechanical, thermal, and integration constraints. 
The key mechanical parameters are summarized in Table~\ref{tbl:LadderConfiguration}. The principal design requirements are as follows:

\begin{itemize}
    \item \textbf{Limited radial space:} The barrel resides at small radii, $\sim7$–$10$ cm around the beam pipe, so the design is constrained by the radial envelope, minimum inter-detector clearance, positioning tolerance, and allowable deflection.
    \item \textbf{Pseudorapidity coverage:}
    The geometry of the ladder must provide coverage of \(|\eta|<1.07\) in both barrels for collision vertices within \(\pm10\) cm along the beam axis (z-axis), matching the acceptance of the other tracking subsystems.
    \item \textbf{Azimuthal coverage:} The barrel must support ladder staves in a cylindrical geometry. Each barrel has two sub-layers such that the active sensor regions cover 100\,\% in azimuth ($0\leq\phi<2\pi$).
    \item \textbf{Low material budget:}
    To minimize multiple scattering and photon conversions in the tracking region, the support structure must be lightweight and low-mass, with an average material budget of approximately $1.1\,\%\ (X/X_0)$ per ladder.
    \item \textbf{Integration of services:} The support structure must interface with cooling, dry air/nitrogen, cabling, readout electronics, and allow installation/maintenance in the sPHENIX environment.
    \item \textbf{Assembly, alignment, integration:}
    The barrel maintains the required radial and azimuthal positions of the ladders, end-rings, and service supports during assembly and under mechanical and thermal loads in nominal detector operating conditions.
    \item \textbf{Installation sequence:} The design must allow the detector to be inserted into the TPC bore as an integrated assembly and allow for the MVTX installation into the INTT barrel bore.
    There are also interfaces with service connections and other considerations of the sPHENIX environment.
\end{itemize}

Detailed specifications on these key items are described below.

\begin{table*}[h]
\begin{center}
\caption{Geometrical and operational parameters of the INTT barrel layers and their radially staggered sub-layers. The pseudorapidity ($|\eta|$) and azimuthal ($\phi$) coverage indicate the active region of silicon sensors.
\label{tbl:LadderConfiguration}}
\begin{tabular}{l|c|c|c|c|c|c|c}
\hline\hline
Barrel         & Quantity & Radius & $|\eta|$ coverage & $\phi$ coverage & $\phi$ overlap  & Clearance  & Avg $X/X_0$ \\
- Sub-layer    &          & (mm)   &                   & (\%)             & (\%)           & (mm)       &  (\%) \\
\hline
Inner barrel   & 24       & -      & -                 & 100              & 2              & 2.00       & 2.15 \\
\hline
- Sub-layer 1a & 12       & 71.90  & 1.37              & 53               & 0              & 0.60       & 1.12 \\
\hline
- Sub-layer 1b & 12       & 78.00  & 1.30              & 49               & 0              & 3.80       & 1.03 \\
\hline
Outer barrel   & 32       & -      & -                 & 100              & 2              & 2.00       & 2.17 \\
\hline
- Sub-layer 2a & 16       & 96.80  & 1.12              & 53               & 0              & 0.60       & 1.12 \\
\hline
- Sub-layer 2b & 16       & 103.30 & 1.07              & 49               & 0              & 3.12       & 1.05\\
\hline\hline
Total          & 56       & -      & 1.07              & 100              & -              & 14.00      & 4.32 \\
\hline\hline
\end{tabular}
\end{center}
\end{table*}

\subsection{Materials}

To minimize mass while maintaining stiffness, the barrel and its service support are fabricated from a combination of lightweight composite and metallic materials:
\begin{itemize}
    \item \textbf{CFC tubes:} The support framework for the service barrel and the barrel structure uses carbon-fiber composite tubes for low mass and high stiffness.
    \item \textbf{Carbon-PEEK stave end-caps:} Some components, such as the stave end-caps, which help to form the barrel structure, use a 30 \% carbon fiber-reinforced PEEK (a high-performance thermoplastic) for its suitable mechanical and radiation-tolerance properties for the expected operation environment.
    \item \textbf{Aluminum end-rings and service brackets:} Aluminum brackets are used for mounting the staves and services. Those are geometrically optimized to minimize mass while maintaining the required stiffness and durable threaded interfaces that would be difficult to implement directly in the composite components.
    \item \textbf{Stainless steel hardware and tube connections:} All hardware is 300 series stainless steel to minimize magnetic permeability and maximize strength, corrosion resistance and thread durability, while still maintaining a reasonable material budget.
\end{itemize}

\subsection{Barrel Geometry and Layer Arrangement}
\label{barrel_and_layer}
\begin{figure} [htb]
    \centering
    \includegraphics[width=0.48\textwidth]{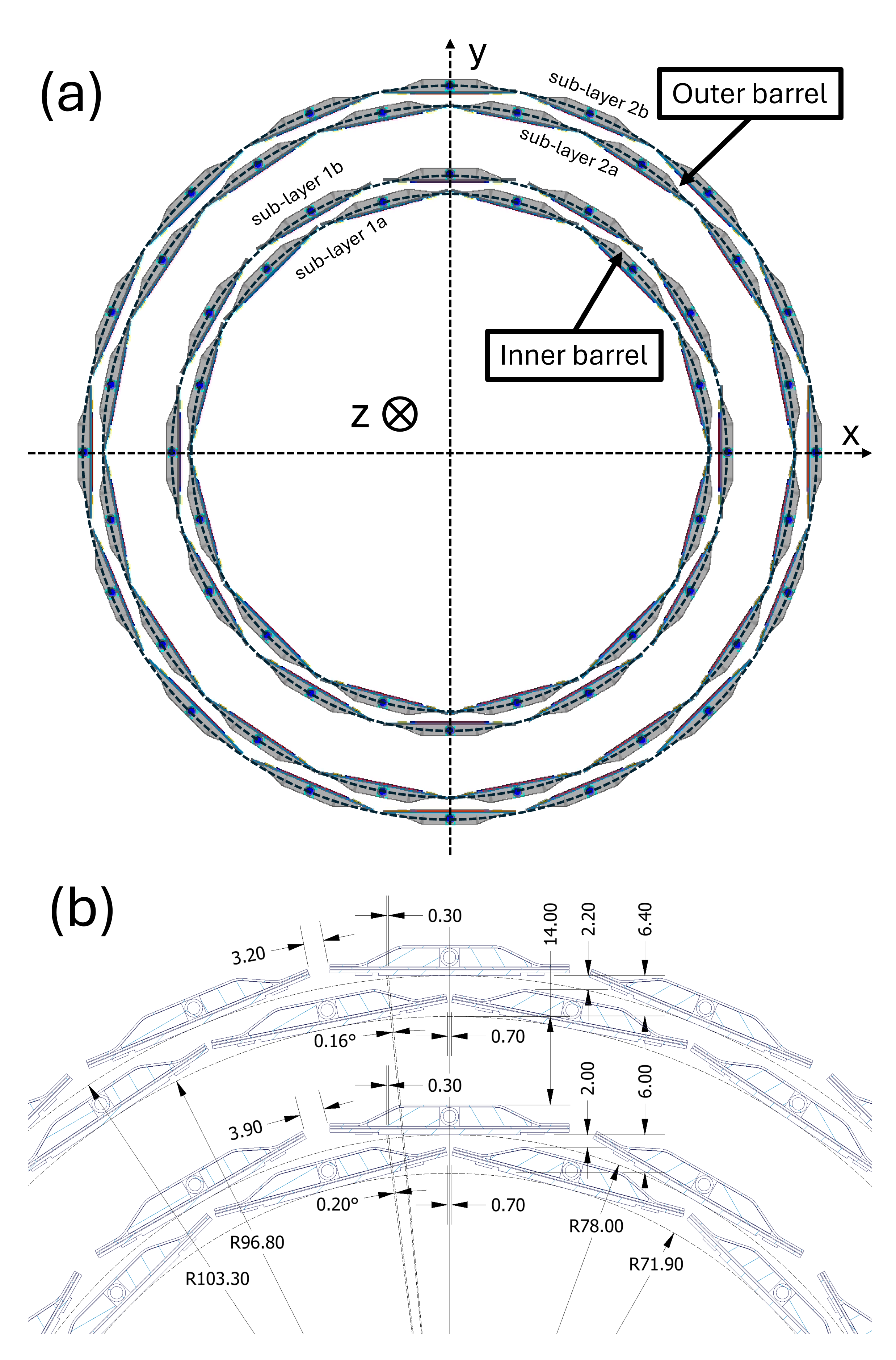}
    \caption{(a) Mechanical cross-sectional drawing of the INTT ladders, viewed along the beam direction, showing the inner and outer barrels. Each barrel consists of two sub-layers in a staggered configuration. (b) Detail view showing the radial position of, and clearance between, the ladders. Values are shown in mm. }
    \label{fig:LaddersXYGeom}
\end{figure}

As shown in Fig.~\ref{fig:LaddersXYGeom}, the INTT barrel consists of two tracking barrels (inner and outer barrel), and each barrel comprises two sub-layers (inner and outer sub-layer) of silicon ladders arranged around the beam pipe.
The ladders are $\sim$50 cm long in all layers, of which 45.6 cm is the sensor area, covering $|\eta|<1.07$ for interaction vertices within $\pm10$\,cm along the z-axis.
A total of 56 silicon ladders are used to construct the sub-layers as follows:
\begin{itemize}
     \item \textbf{Sub-layer 1a:} Inner sub-layer of the inner barrel at a radius of approximately 71.9\,mm, containing 12 ladders.
     \item \textbf{Sub-layer 1b:} Outer sub-layer of the inner barrel at a radius of approximately 78.0\,mm, containing 12 ladders.
     \item \textbf{Sub-layer 2a:} Inner sub-layer of the outer barrel at a radius of approximately 96.8\,mm, containing 16 ladders.
     \item \textbf{Sub-layer 2b:} Outer sub-layer of the outer barrel at a radius of approximately 103.3\,mm, containing 16 ladders.
\end{itemize}
The inner and outer sub-layers are arranged tangentially and staggered in both radius and azimuth ($\phi$) so that the active sensor regions cover 53\,\% and 49\,\% in azimuth, respectively, together achieving 100\,\% azimuthal coverage with a 2\,\% overlap region, which corresponds to 0.3\,mm on each sensor ($\sim$4 silicon strip channels).
Although the design offers complete azimuthal coverage for charged particles traveling in straight trajectories without a magnetic field, particles with curved trajectories in a magnetic field may escape through the gaps between sensors. As shown in Fig.~\ref{fig:LaddersXYGeom} (b), the minimum gaps between sensors on inner and outer sub-layers are 6.0\,mm and 6.4\,mm for inner and outer barrels, respectively.

Mechanically, the end-rings hold the ladders at the correct radius, ensure azimuthal spacing and tangential arrangement, and permit ladder mounts and service attachments.
Therefore, the end-rings provide distinct mounting surfaces at the four required radii and azimuthal offsets.
The design of the aluminum end-rings is critical, providing the structural anchor for ladders and transferring loads to the support tubes and ultimately the detector cradle through the INTT support structure.
Most dimensions have a prescribed +/-0.125\,mm tolerance with the stave mounting faces having a +0.000/-0.125\,mm tolerance and mounting holes having a +/-0.05\,mm tolerance.

\subsection{Detector Assembly}

\begin{figure*} [ht]
    \centering
    \includegraphics[width=1.0\textwidth]{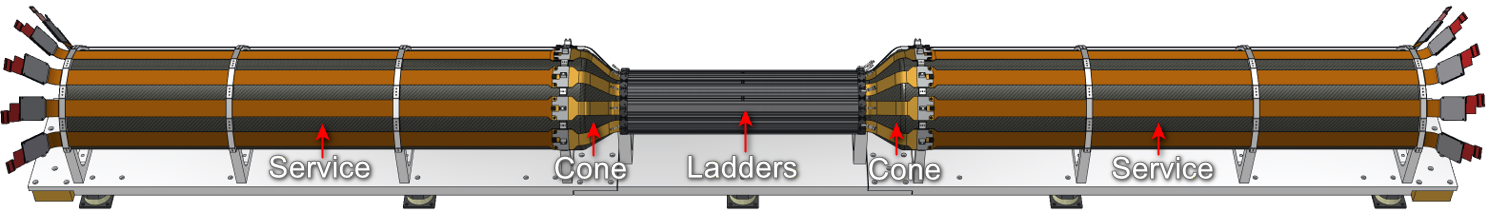}
    \caption{First sub-layer installation, showing the fixture, ladders, cones, and services.}
    \label{fig:BarrelAsm}
\end{figure*}

For assembly and installation, each INTT barrel is divided into two identical, rotationally symmetric half-barrels. 
A dedicated fixture was designed to support the ladders and service barrel during assembly.
The half-barrels are then assembled from the inside out so that the inner radius rests on the fixture.
The fixtures are mounted on two vertically adjustable carts, with vibration dampers, to allow for easy access and maneuverability during assembly and transportation. See Fig.~\ref{fig:BarrelAsm} for a partially assembled half-barrel on the assembly fixtures.

\begin{enumerate}
    \item The assembly begins by placing the carbon fiber inner service barrel tubes, cones and skin onto the fixture, immediately followed by the innermost aluminum end-rings and service brackets. The hardware that attaches the brackets to the tubes and the temporary hardware attaching the fixture to the brackets can be installed from the inner radius.
    \item The first ladder for the first sub-layer (1a) can then be installed, attaching the stave end-caps to the flats on both end-rings, followed by its corresponding services to the service brackets. This is repeated for all ladders for that sub-layer.
    \item After the first sub-layer is completed, the next layer of service brackets is installed on top of the previously installed set.
    \item The second sub-layer (1b) of ladders can be installed directly on top of the previous, followed by its corresponding services to the service brackets.
    \item After the second sub-layer is completed, the next layer of service brackets is installed on top of the previously installed set, and the next end-ring is installed over the end-caps of the last layer of ladders.
    \item Steps 2-5 are repeated for the next two sub-layers (2a and 2b).
    \item Then the outer skin, cones and service barrel tube is installed over and attached to the aluminum end-rings and service brackets. On the exterior of the outer service barrel tube the ball transfer brackets that allow for installation are attached, completing the half-barrels.
\end{enumerate}

\begin{figure} [htb]
    \centering
    \includegraphics[width=0.48\textwidth]{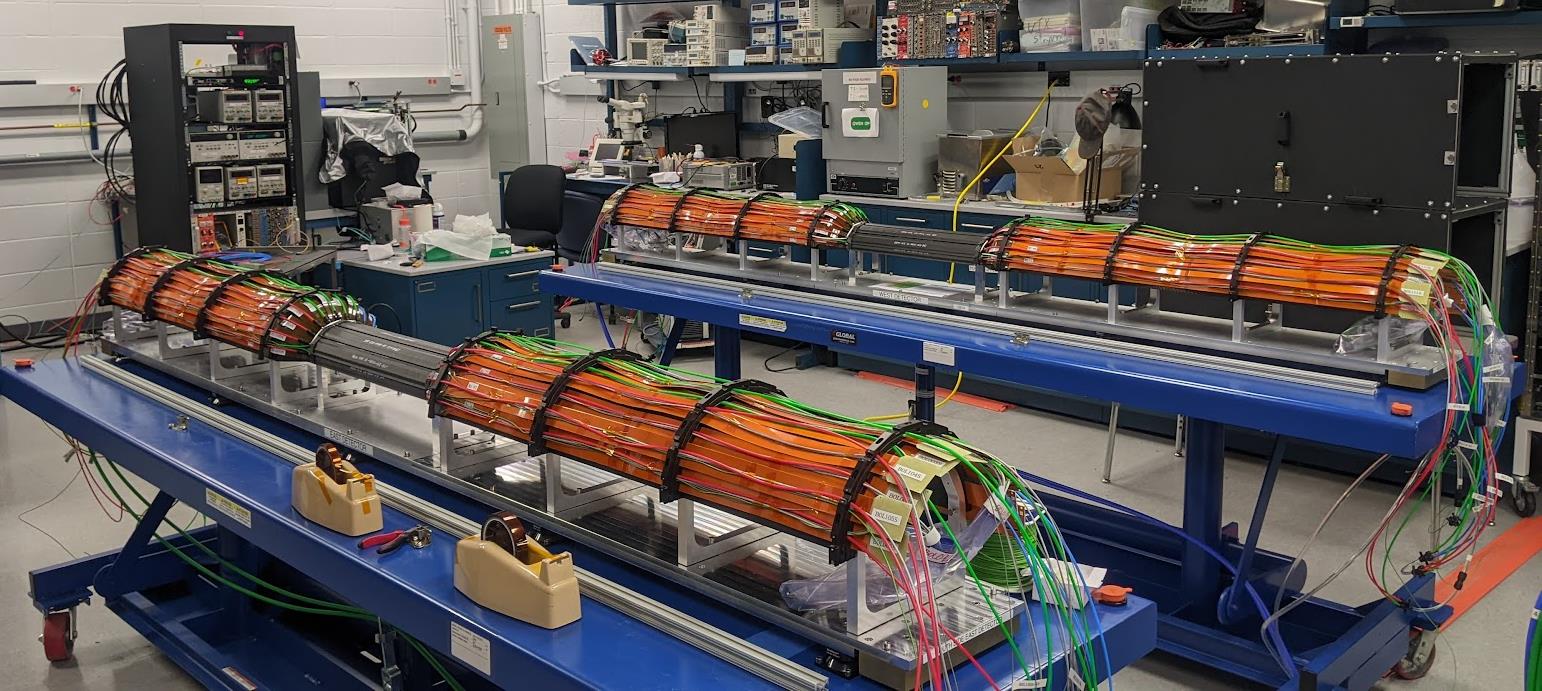}
    \caption{Fully assembled INTT half-barrels at BNL Physics Department silicon laboratory.}
    \label{fig:HalfBarrels}
\end{figure}

\begin{figure} [htb]
    \centering
    \includegraphics[width=0.48\textwidth]{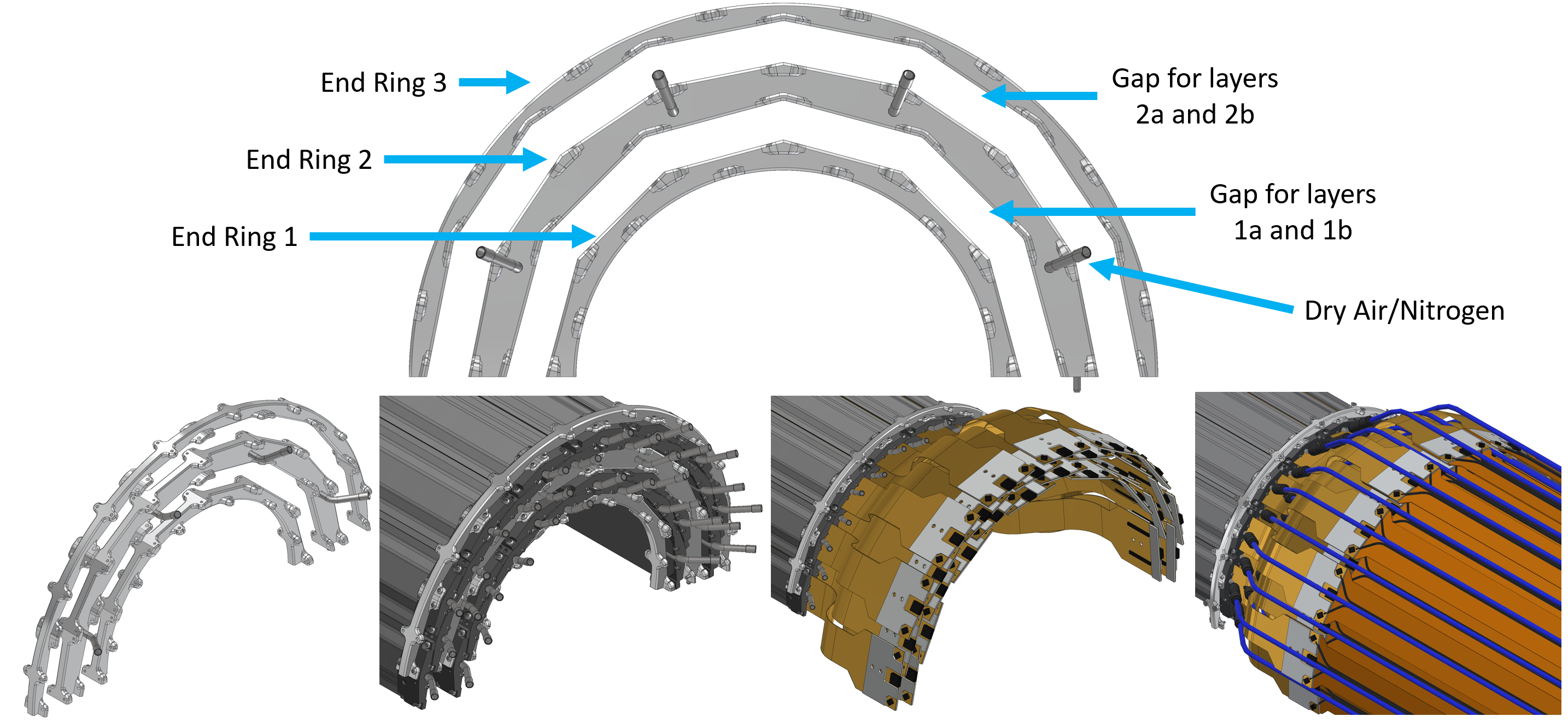}
    \caption{End-ring configuration with four levels of detail of the ladders and their services.}
    \label{fig:EndRings}
\end{figure}

As shown in Fig.~\ref{fig:HalfBarrels}, each INTT half-barrel is fully assembled and tested in the Physics Department silicon laboratory at BNL. The two halves are then assembled together in the lab for testing to verify mechanical fit and functional performance. 
The achieved mechanical fit is evaluated  with collision data in Section~\ref{sec_INTT_acceptance}.
The final integration of the two half-barrels occurs around the beam pipe within the sPHENIX interaction region, inside the inner bore of the TPC. See Fig.~\ref{fig:EndRings} for different levels of detail for the end-rings, ladders, and services.

\subsection{Support, Mounting, Installation and Integration}
\begin{figure} [htb]
    \centering
    \includegraphics[width=0.48\textwidth]{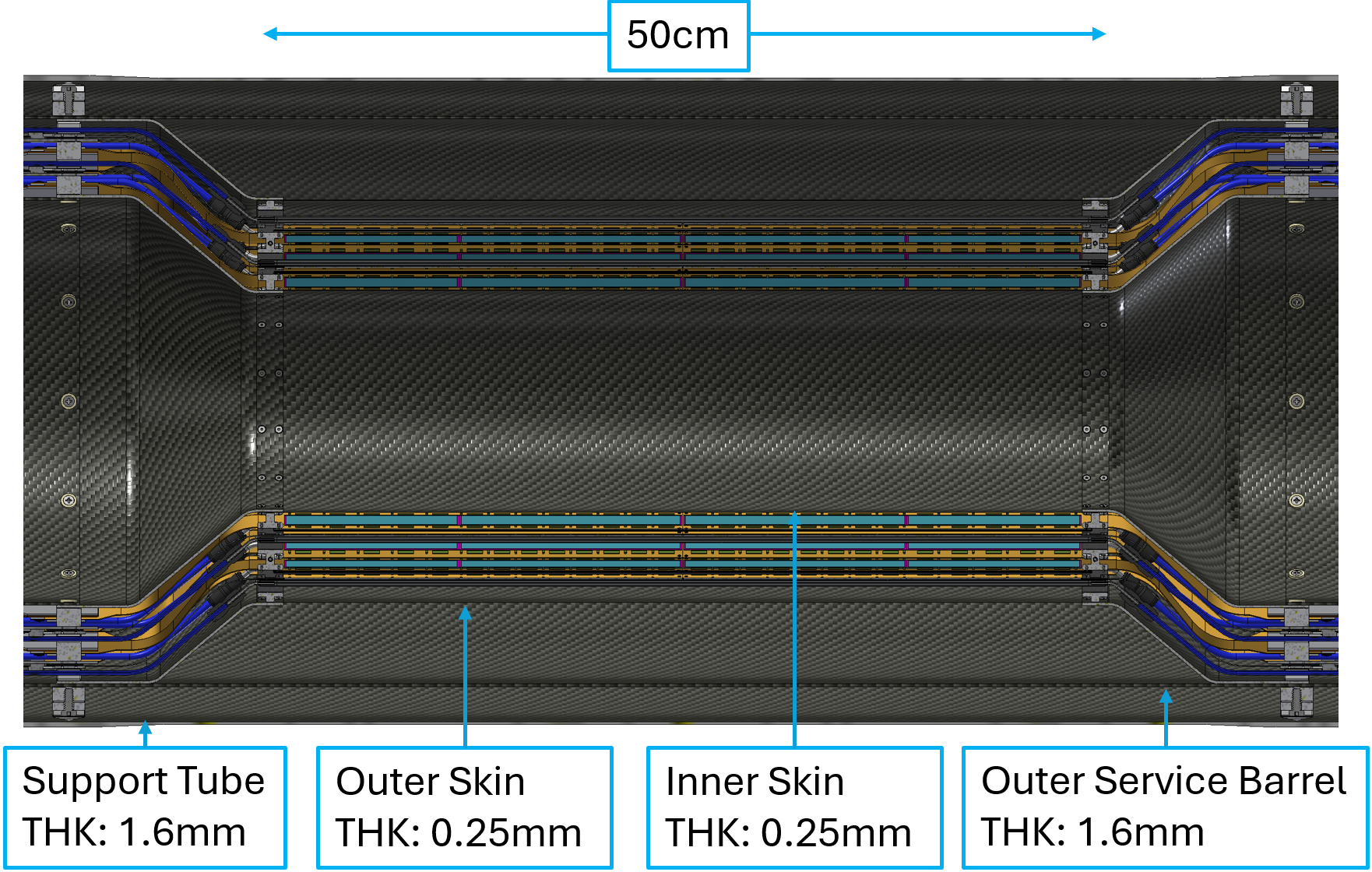}
    \caption{Central carbon fiber structures, showing from the inside out the: inner skin, outer skin, outer service barrel and support tube.}
    \label{fig:CentralStructure}
\end{figure}

In addition to ladders, the central barrel structure consists of four layers of carbon fiber as described in the assembly: two skins that are both 0.25 mm thick ($0.10\,\%\,X/X_0$) and the outer service barrel and the support tube that are both 1.6 mm thick ($0.62\,\%\,X/X_0$) as shown in Fig.~\ref{fig:CentralStructure}. The skins serve primarily as protection from incidental contact, debris, moisture infiltration and electrical noise.
The outer service barrel stiffens the detector structure over its full length, while the support tube forms part of the INTT barrel support structure.

\begin{figure} [htb]
    \centering
    \includegraphics[width=0.45\textwidth]{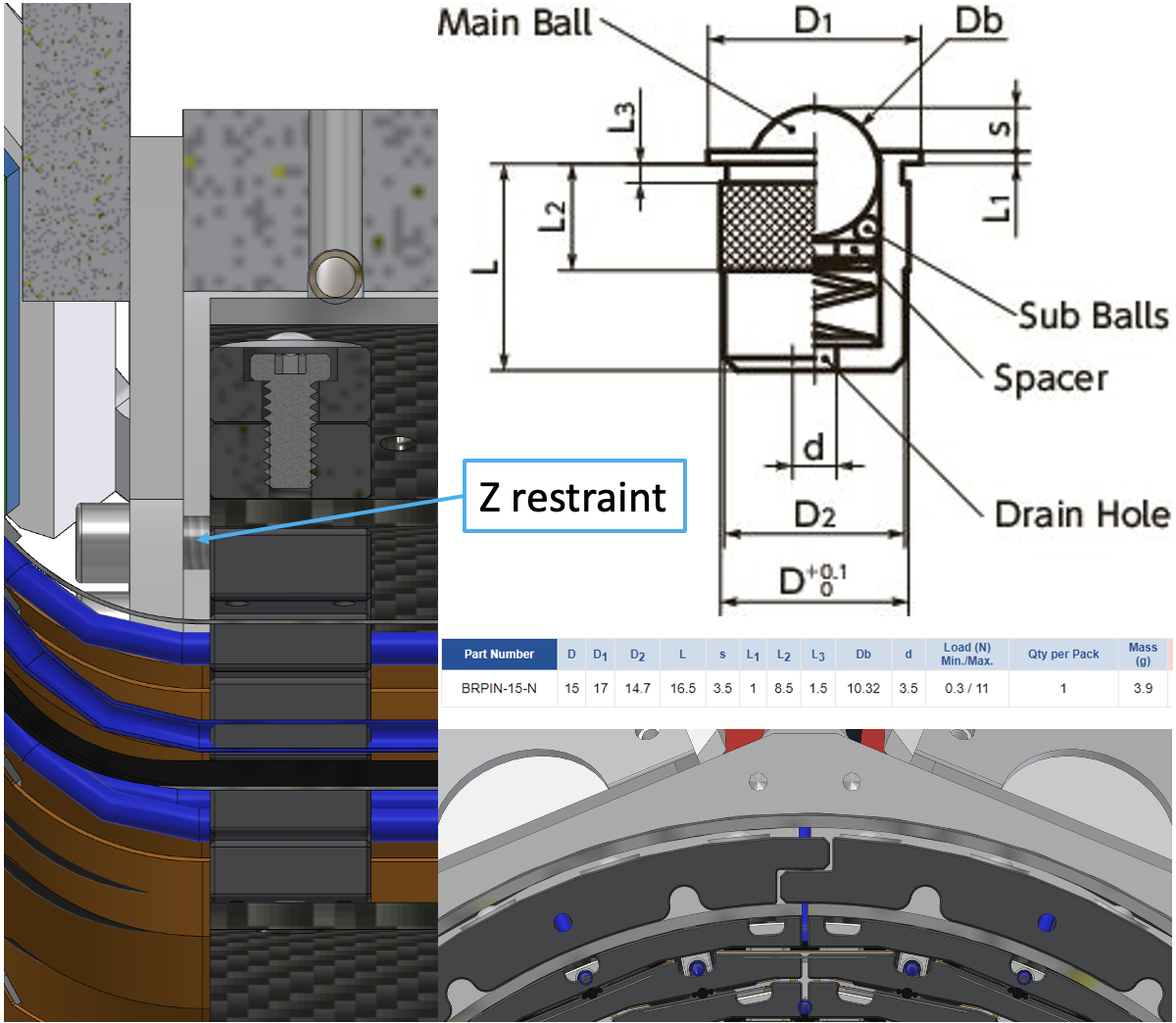}
    \caption{The ball transfers and mounting brackets on the exterior of the outer service barrel.}
    \label{fig:BallTransfers}
\end{figure}

The INTT barrel support structure consists of a full carbon fiber tube, 375 mm in diameter, spanning the length of the TPC (approximately 2.75 m) with aluminum rings and I-beams that attach to the inner diameter of the iHCal. The structure features a smooth surface on the tube inner diameter to allow for the ball transfers to slide during installation, a thick-walled tube except for the region close to the IP, sizable I-beams to minimize deflection, and leveling feet to precisely align the detector position.
The INTT is then assembled around the beam pipe in a clamshell configuration at a position offset  from the interaction point (IP) along the z-axis, and installed by transferring it from the installation fixture onto the support tube using the exterior ball transfers, as shown in Fig.~\ref{fig:BallTransfers}. The MVTX is subsequently installed in a similar manner inside the INTT.

%%%%%%%%%%%%%%%%%%%%%%%%%%%%%%%%%%%%%%%%%%%%%%%%%%
%%%%%%%%%%%%%%%% LV/HV System %%%%%%%%%%%%%%%%%%%%
%%%%%%%%%%%%%%% Akitomo, Ivan %%%%%%%%%%%%%%%%%%%%
%%%%%%%%%%%%%%%%%%%%%%%%%%%%%%%%%%%%%%%%%%%%%%%%%%
\section{Power Systems}
\label{Power system}
The INTT power system includes two racks positioned on the north and south sides of the interaction region. The south (or north) rack supplies power to the south (or north) half silicon ladders. Each rack contains the bias voltage modules with noise filters for the silicon sensors and the low voltage (LV) modules for the ROC cards, beam clock distribution boards, and silicon ladders' FPHX chips.

\subsection{Bias voltage system}
The applied bias voltage for full depletion of each silicon sensor is 100\,V. This voltage is supplied via the WIENER MPOD module (model EHS F605p-F). The INTT detector is composed of 112 half-ladders, with each half-ladder featuring two input channels for bias voltage, dedicated to type-A and type-B silicon sensors, thus providing a total of 224 bias voltage input channels. Each MPOD module features 16 output channels that connect to a ROC via 12-meter-long co-axial cables and is capable of supplying up to 8 half-ladders' bias voltages. Together, 16 MPOD modules are utilized, with 8 for each of the south and north racks, for a total of 16 ROCs. For the ultimate INTT setup, seven half-ladders are linked to one ROC, and fourteen of the sixteen MPOD channels are utilized with two spare channels. Control of the bias voltage, including on/off, voltage setting, and ramping speed setting, is carried out using a newly created SNMP/C++-based software with a Python GUI. Typical leakage currents for the input channel fall between 5 to 10~$\mu$A. Leakage current readings for each channel are recorded in the PSQL database at a rate of one measurement per second, and we use Grafana~\cite{Grafana} for monitoring the detector's operational status in real time.

\subsection{Low Voltage System}
\begin{figure} [htb]
    \centering
    \includegraphics[width=0.48\textwidth]{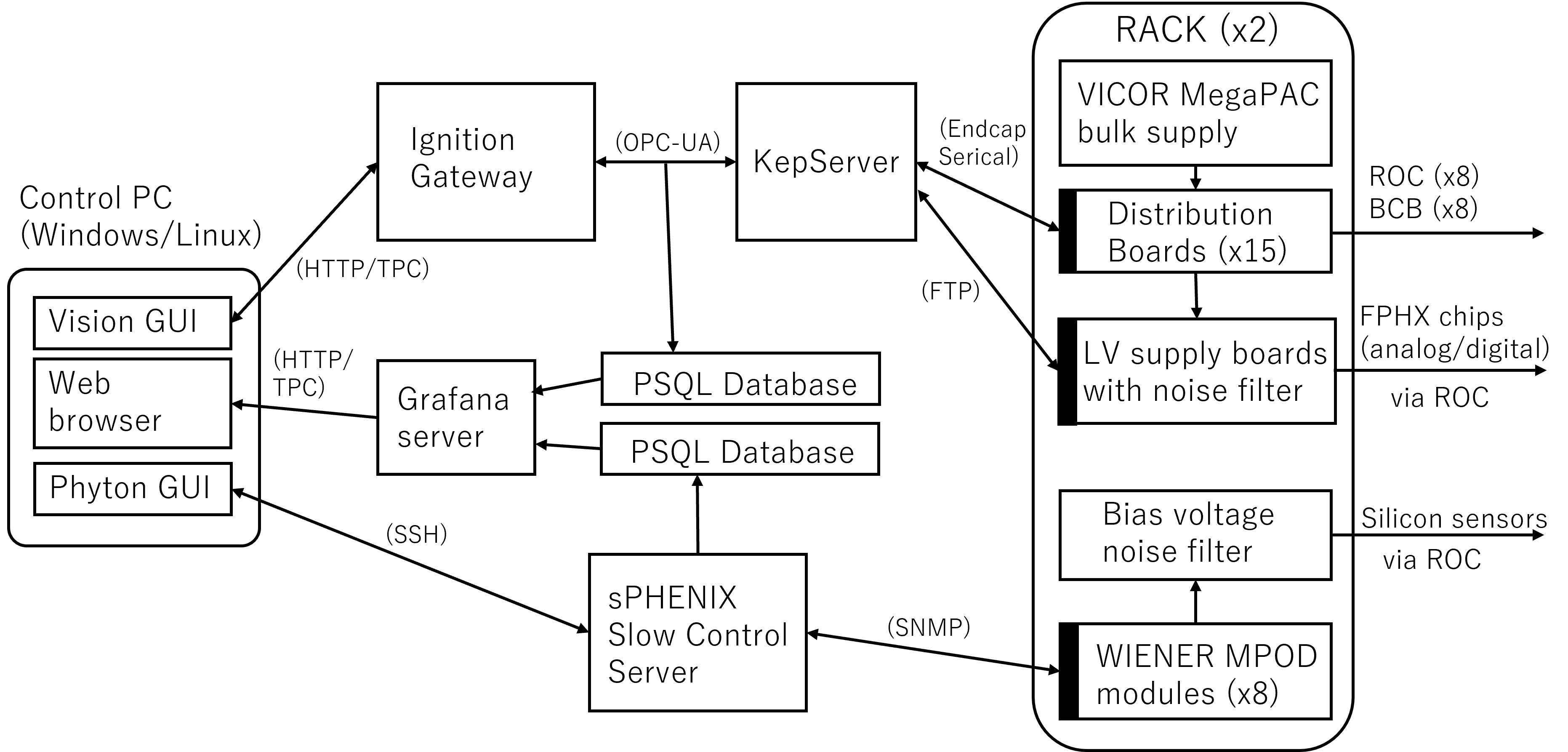}
    \caption{A diagram of the slow control for the INTT power system.}
    \label{fig:SCDiagram}
\end{figure}

The VICOR MegaPAC bulk supply, with its distribution boards, powers the ROC, the beam clock boards, and the silicon ladders' FPHX chips, according to the INTT LV specifications. The FPHX chips receive power application through noise filter boards. For the INTT LV system control, the Ignition framework is being utilized, as depicted in Fig.~\ref{fig:SCDiagram}. The control and real-time monitoring of voltages/currents, implemented with the Ignition VISION module~\cite{Ignition}, are handled by the operator's GUIs, while Grafana with a PSQL database is used for long-term monitoring.

%%%%%%%%%%%%%%%%%%%%%%%%%%%%%%%%%%%%%%%%%%%%%%%%%%
%%%%%%%%%%%%%%%%% ClockSystem %%%%%%%%%%%%%%%%%%%%
%%%%%%%%%%%%%%%%% Itaru       %%%%%%%%%%%%%%%%%%%%
%%%%%%%%%%%%%%%%%%%%%%%%%%%%%%%%%%%%%%%%%%%%%%%%%%
%%%%%%%%%%%%%%%%%%%%%%%%%%%%%%%%%%%%%%%%%%%%%%%%%%%%%%%%%%%%%%%%%%
\section{Beam Clock Distribution System}
\begin{figure}[htb]
    \centering
    \includegraphics[width=0.45\textwidth]{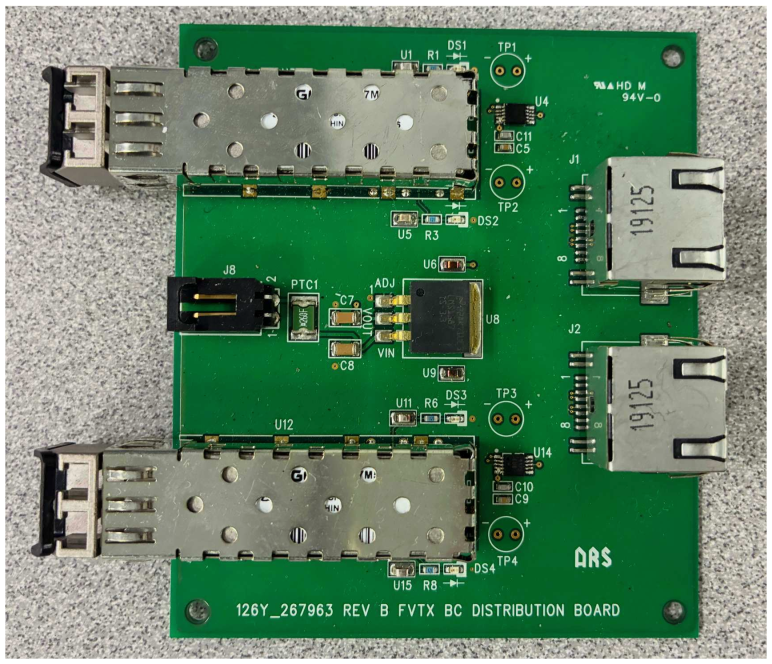}
    \caption{The top view of the beam clock board.}
    \label{fig:BeamClockBoard_photo}
\end{figure}

%%%%%%%%%%%%%%%%%%%%%%%%%%%%%%%%%%%%%%%%%%%%%%%%%%%%%%%%%%%%%%%%%%
\label{clock_board}
The beam clock board (BCB) for the INTT detector provides low-jitter, phase-aligned clock and timing signals to the ROC. The board receives the 9.38 MHz global beam clock and Start clock signals from the global timing module (GTM) via FELIX servers through an optical fiber, and converts them into LVDS signals. The function of the Start clock signal will be discussed in Section~\ref{Felix}. These signals are transmitted to the ROC via an RJ45 modular connector. A photograph of the BCB is shown in Fig.~\ref{fig:BeamClockBoard_photo}. 

The BCB is a simplified derivative of the beam clock distribution board (BCDB) developed for the FVTX detector~\cite{Aidala:2013vna}. In contrast to the BCDB, the BCB omits the fan-out functionality and is dedicated to optical-to-electrical signal conversion. In addition, the output interface was changed from a Molex 8-pin connector (model 87831-0820) to an RJ45 modular connector. The BCDB distributed clock signals to up to seven ROCs, whereas the BCB serves a single ROC.
The principal components of the BCB are summarized in Table~\ref{tbl:BCB_Specification}. The LVDS splitter (MAX9175) operates up to 670 MHz and provides low skew and low jitter performance. The specified random jitter is 1.0 ps (RMS, maximum), which is suitable for high-speed timing applications.

\begin{table}[h]
\begin{center}
\caption{Specification of the beam clock board.}
\label{tbl:BCB_Specification}
\begin{tabular}{l|c|c}
\hline\hline
 Item             & Make    & Model \\
\hline
 Optical Transceiver   & AVAGO   & AFBR-57M5APZ \\
 LVDS Splitter         & Maxim   & MAX9175EUB+T \\
 Modular Jack          & TE Connectivity & 6339191-5 \\
\hline\hline
\end{tabular}
\end{center}
\end{table}

During operation of the FVTX detector, rare and unexpected resets of the ROC were observed.
Although its origin was not conclusively identified, transmission of clock signals over twisted-pair cables longer than 1\,m was suspected as a possible source of instability.
In the BCDB configuration, such cable lengths were required to distribute clock signals to multiple ROCs located at different positions.
Based on this experience, the BCB design assigns one board per ROC and minimizes the cable length between the BCB and ROC. The BCB is installed directly behind the ROC backplane, reducing the transmission distance.

Custom beam clock cables were fabricated using Category 5e LAN cables with an RJ45 modular plug on one end and a Molex 51110-0851 plug on the other, compatible with the ROC input. The cable length is 15 cm to suppress signal degradation. The specification of the beam clock cable is summarized in Table~\ref{tbl:BCLK_cable}.

\begin{table}[h]
\begin{center}
\caption{Specification of the beam clock cable.}
\label{tbl:BCLK_cable}
\begin{tabular}{l|c}
\hline\hline
 Cable            & LAN \\
 Category         & 5e \\
 Connector 1      & Modular plug RJ45 \\
 Connector 2      & 51110-0851 (Molex) \\
 Length           & 15 cm\\
\hline\hline
\end{tabular}
\end{center}
\end{table}

No packet loss has been observed during three years of INTT operation, demonstrating the stability of the present configuration.

%%%%%%%%%%%%%%%%%%%%%%%%%%%%%%%%%%%%%%%%%%%%%%%%%%
%%%%%%%%%%%%%%%% INTT DAQ   %%%%%%%%%%%%%%%%%%%%%%
%%%%%%%%%%% Itaru, Takashi, Raul %%%%%%%%%%%%%%%%%
%%%%%%%%%%%%%%%%%%%%%%%%%%%%%%%%%%%%%%%%%%%%%%%%%%
\section{Data Acquisition System}

\label{INTT DAQ}
This section describes the data acquisition (DAQ) chain and readout electronics involved in reading out the INTT barrel detector including 56 silicon ladders.
One silicon module has two silicon strip sensors mounted on an HDI that houses 26 FPHX ASIC chips that read out 128 silicon strip channels each. The total number of channels in each silicon module is $26\times 128$, which is equivalent to 3,328 channels per silicon module, leading to 372,736 in total for the INTT.
Each silicon module is connected to one ROC through the readout chain, consisting of the HDI cable, the BEX, and the CC.
The detailed specification of the silicon ladder module is described in \cite{Nakagawa2025}.

\begin{figure*}[h]
    \centering
    \includegraphics[width=0.9\textwidth]{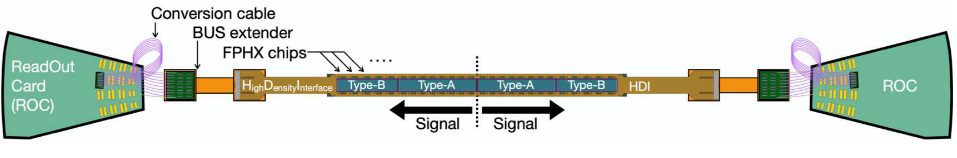}
    \caption{Schematics of readout electronics from silicon module to ROC}
    \label{readout_cable_schm}
\end{figure*}

\begin{figure}[htb]
\begin{center}
\includegraphics[width=0.4\textwidth]{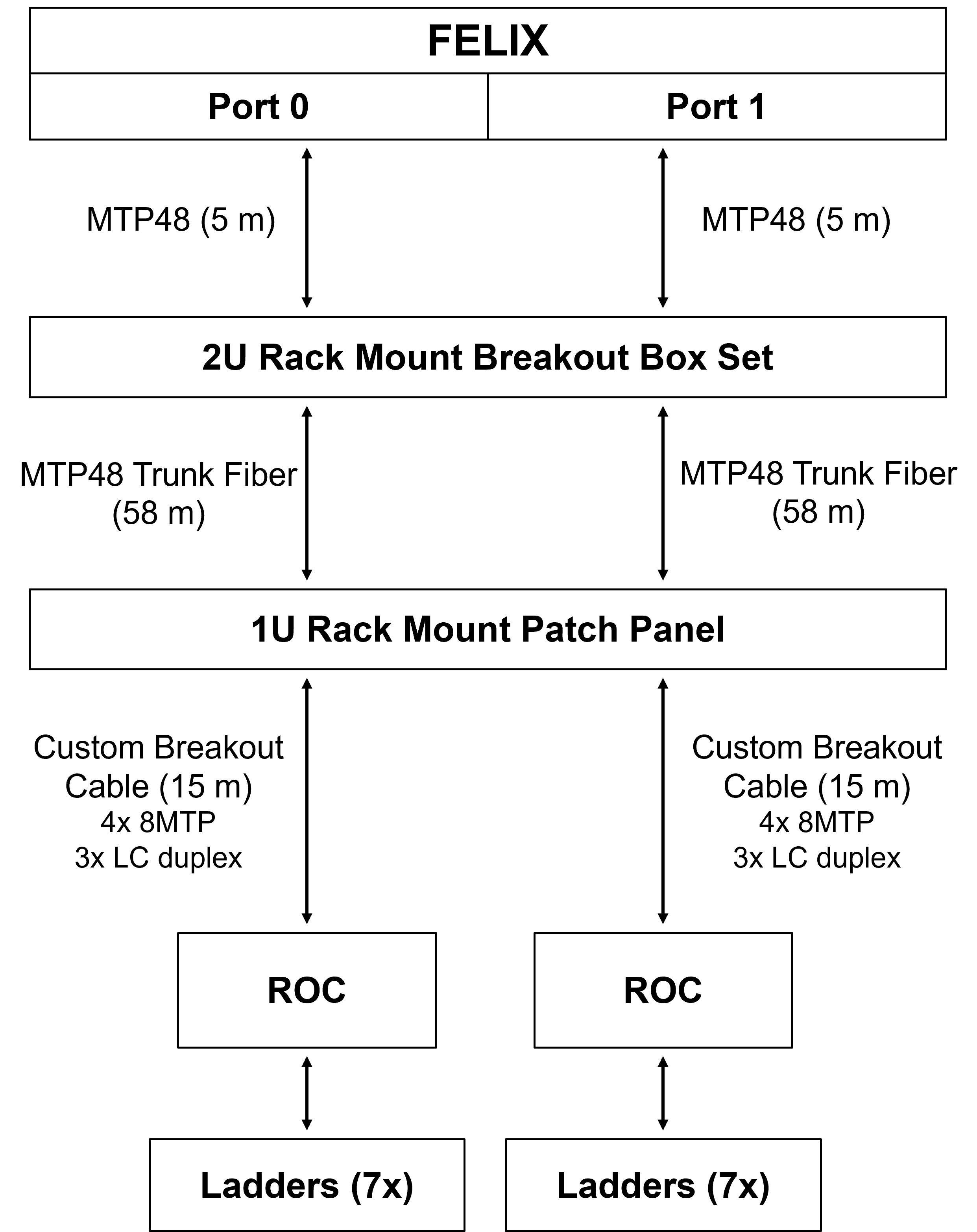}
\caption{
A schematic diagram of fiber distribution.
\label{fig:FelixFiberDiagram}
}
\end{center}
\end{figure}

The ROC is a circuit board designed originally for the FVTX detector~\cite{Aidala:2013vna} in the PHENIX experiment and reused in INTT. The role of the ROC in INTT is to control, power, and collect data from silicon modules.
It reformats the data sent by the FPHX chips and retransmits them downstream in the DAQ chain. It was designed with a trigger-less, data-push architecture, which, unlike triggered systems, does not store, select or sort information before sending it downstream. The ROC board comprises 16 pairs of connectors, each referred to as a ROC port.
Of these 16 ROC ports, 12 are compatible with the INTT silicon modules.
Due to the INTT barrel configuration, only 7 of the 12 compatible ROC ports are used during the normal operation; the remaining 5 ports are left unterminated and serve as spares.
Furthermore, the FELIX board data aggregation module and the ROC are linked through optical fibers, as summarized in Fig.~\ref{fig:FelixFiberDiagram} and Table~\ref{tbl:FELIXFibers}.
The board also has an on-board calibration system for the FPHX chips. It delivers to the chips a controlled and low-noise voltage pulse, synchronous to the beam counter orbit (BCO) which refers to the unique identifier for a beam crossing. Amplitude, frequency, and number of pulses are configurable, allowing a complete scan of every silicon sensor channel and ADC~\cite{Aidala:2013vna}. For INTT operations, the firmware of the ROC was preserved, and the hardware was mostly preserved, minus refurbishing and component updates. Notably, to account for a larger voltage drop in the cables connecting the ROC to the silicon modules, footprint-compatible regulators with a higher regulated voltage were used~\cite{Nakagawa2025}.

\begin{table}[h]
\begin{center}
\caption{FELIX optical channel allocation per port
\label{tbl:FELIXFibers}}
\begin{tabular}{l|c|c|c}
\hline\hline
                & Component             &  Qty          & Link Rate \\
\hline
                & Slow Control          & 1 Duplex      & 2.1 Gbps \\
MTP48           & Data Link             & 14 Simplex    & 2.5 Gbps \\
                & Beam Clock            & 1 Simplex     & 1.801 Gbps \\
                & Start Clock           & 1 Simplex     & 1.801 Gbps \\
\hline
LC Duplex       & Timing Link     & 1 Duplex             & 1.2 Gbps\\
\hline\hline
\end{tabular}
\end{center}
\end{table}

%%%%%%%%%%%%%%%%%%%%%%%%%%%%%%%%%%%%%%%%%%%%%%%%%%%%%%%%%%%%%%%%%%%%%
\subsection{FELIX Server}
\label{Felix}

\begin{figure}[htb]
\begin{center}
\includegraphics[width=0.48\textwidth]{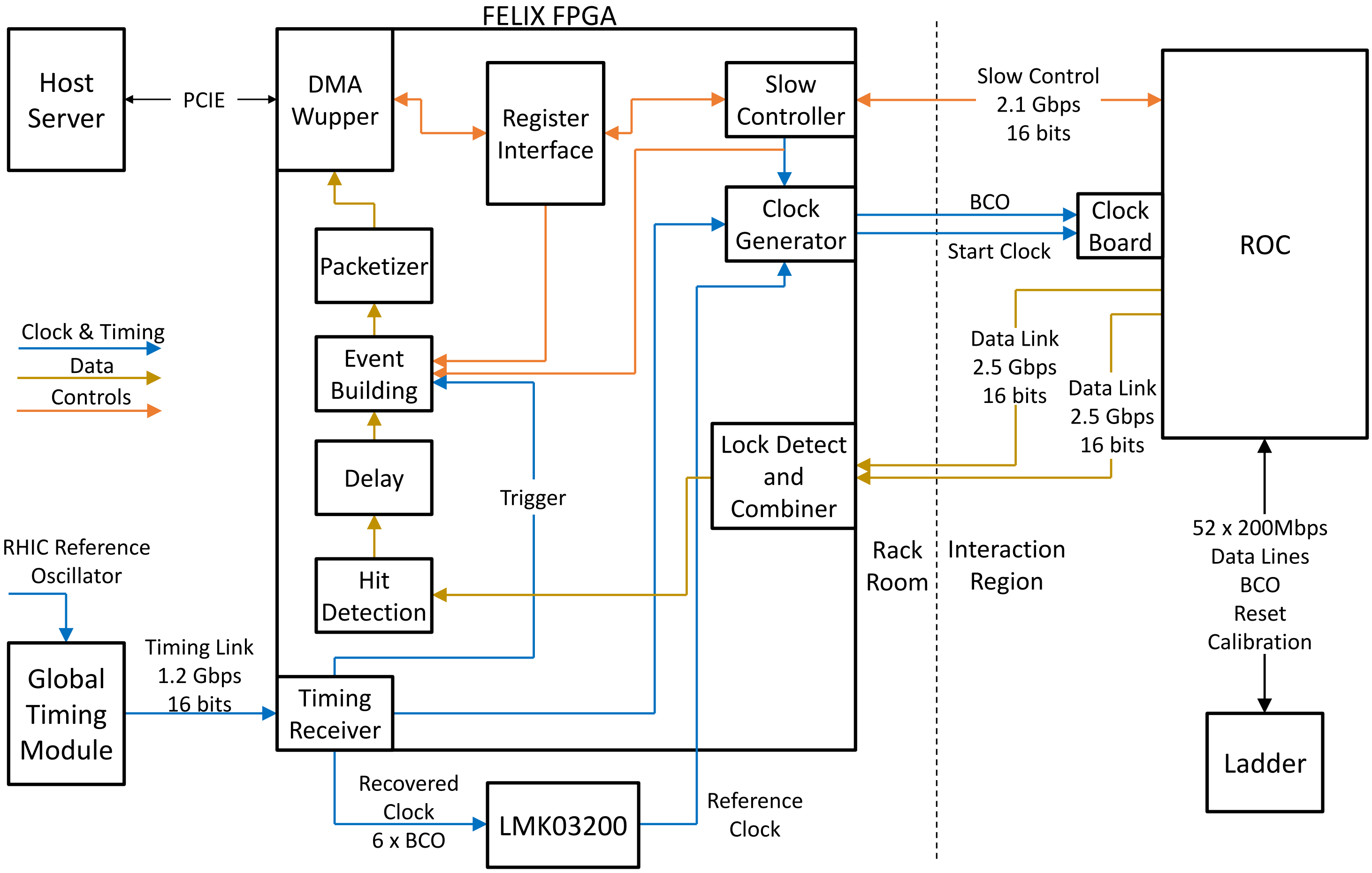}
\caption{
A schematic diagram of the FELIX-based data acquisition.
\label{fig:FelixFlowChart}
}
\end{center}
\end{figure}

The FELIX server is a commercial 4U AMD EPYC 7413 server with 2 24-core processors. It hosts the FELIX board~\cite{Anderson_2016} which is a custom-designed PCIe card with a 16-lane PCIe Gen3 interface originally designed by the ATLAS TDAQ Collaboration~\cite{Felix}.
With 8 Avago MiniPOD transceivers, it totals 48 duplex optical fiber channels with each optical link's bandwidth of 10+ Gbps. The board contains a Xilinx Ultrascale FPGA XCKU115. It functions as a data aggregation module, concentrating data, providing clocks and synchronizing RHIC's fast and slow controls to the ROCs. It interfaces directly with 2 ROCs and their clock boards, the GTM, and a server through PCIe, as shown in Fig.~\ref{fig:FelixFlowChart}. 
The control of the FELIX through the PCIe interface, as well as DMA buffer management, is handled by a Linux kernel driver created at BNL. Below are some definitions of relevant readout and control parameters and concepts:

\begin{itemize}
    \item \textbf{Open time: } The amount of time, in units of bunch crossings, during which hits will be accepted after a trigger or strobe is received by the FELIX.
    \item \textbf{Number of crossings: } Used to filter the hits coming into the FELIX by their timestamp, as assigned by the FPHX chip.
    \item \textbf{Strobe: } With a similar function as the trigger bit received by the FELIX, the strobe starts a data acquisition window when in streaming mode. It is sent by the GTM during RHIC's abort gap.
    \item \textbf{BCO frequency: } The frequency of the BCO oscillator is a 9.38 MHz clock that is sent to the FPHX chips through the ROC, FELIX, and GTM.
    \item \textbf{Start clock: } A clock generated by the FELIX and edge-aligned with the BCO. Its purpose is to send synchronous commands to the ROC, which executes them with fixed latency. It is used to start calibration and to reset the timestamp of the FPHX chips.
    \item \textbf{Scheduler: } List of commands the GTM sends to the FELIXes at a predetermined time.
    \item \textbf{Event: } An event comprises all hits associated with a strobe (in the streaming readout context) or trigger (in the triggered readout context), as well as the 40-bit BCO counter transmitted by the GTM.
\end{itemize}

\subsubsection{Detector Synchronization}
Each FELIX board is connected to the GTM via optical link, as seen in Fig.~\ref{fig:FelixFlowChart}. The GTM is a central part of sPHENIX's timing and control. It distributes to the FELIXes: RHIC's reference clock; commands synchronous to specific RHIC bunches; operation mode flags; triggers and strobes; and a 40-bit BCO counter common to all detectors. It allows all FELIXes to be synchronized among themselves and with other sPHENIX subsystems. 

The FELIX board recovers the collider clock from the GTM-FELIX optical link by operating its GTM transceiver in fixed-latency mode, which enables clock recovery with a lower clock phase uncertainty with respect to the GTM reference. The recovered clock is, then, fed to an external PLL (LMK03200) that is used to drive the FELIX GTM that sends the BCO and start clock through fiber links to the clock board as described in Section~\ref{clock_board}. Overall, the phase between two different clocks sent to any two ROCs is no greater than 21 ns.

\subsubsection{Calibration Mode}
The calibration mode is controlled by the ROCs, which inject pulses of varying amplitudes into each of the FPHX ASIC front-end channels. Once the FELIX sends a calibration request to the ROC, it starts a routine that sends a predetermined number of pulses to a single channel of every chip, allowing evaluation of sensor integrity and baseline noise levels.

\subsubsection{Triggered Readout Mode}
The triggered mode comprises the selection of hits from the crossing of interest by their timestamps, as assigned by the FPHX chips. The FELIX maintains a shadow BCO counter that mirrors the timestamp counters in the FPHX chips. This mode is also able to collect hits for up to \SI{13}{\micro\second} after the triggered crossing.
The number of crossings contained by a single trigger event is defined by the \textit{open time} and \textit{number of crossings} parameters.
The triggered operation, from a FELIX firmware perspective, assumes a hit will reach the Event Building module after a trigger is received. To guarantee such a condition, an artificial delay line is inserted in the data path, as shown in Fig.~\ref{fig:FelixFlowChart}. When a trigger is received by the FELIX, an event is created, and the time window during which hits will be accepted as part of such an event starts. As hit information comes in from the ROCs, they are filtered by their timestamps, and are saved as part of the ongoing event. Once the window times out, the event is finalized, and no more hits are accepted.

\subsubsection{Streaming Readout Mode}
The streaming readout mode captures all hit information transmitted by the ROCs, regardless of their timestamps. Its operation mode involves resetting the FPHX chips' timestamps and issuing a strobe during the abort gap. Once a strobe is received by the Event Building module, a streaming event starts.

In streaming mode, the GTM sends a strobe to the FELIXes during the RHIC abort gap, starting an event that captures all the hits related to the incoming RHIC revolution.
To improve data reconstruction and online data monitoring, when in streaming mode, the crossing counters of all FPHX ASICs are synchronously reset, also during the abort gap. That results in the hit timestamps coming from the chips to track RHIC's bunch number.
To mitigate the background contribution from hot channels, the threshold of DAC value in the FPHX chip for the streaming mode is raised about 20\% higher than that of the triggered mode.
% Insert RCDAQ section here. One paragraph, with the tone of conclusion.
\subsection{RCDAQ}
RCDAQ~\cite{RCDAQ} is a data acquisition package used in the PHENIX and sPHENIX experiments. When data from the FELIX is available at the server DMA buffer, RCDAQ reads such buffer and decodes the data to extract events and hit information.
Finally, RCDAQ saves the decoded information on other servers called bufferboxes, which serve as local and temporary storage for the entire sPHENIX data before it's saved on BNL's permanent data storage infrastructure.

%%%%%%%%%%%%%%%%%%%%%%%%%%%%%%%%%%%%%%%%%%%%%%%%%%
%%%%%%%%%%%%%%%% INTT DAQ   %%%%%%%%%%%%%%%%%%%%%%
%%% Ryotaro,Genki,Akitomo, Cheng-Wei, Takashi %%%%
%%%%%%%%%%%%%%%%%%%%%%%%%%%%%%%%%%%%%%%%%%%%%%%%%%
\section{Performance}
\label{BaarrelPerformance}
This section presents performance benchmarks from the INTT, using data collected during the operation of sPHENIX.

\subsection{Live Channel Fraction} %Jaein
\label{sec_live_bad_chan}

\begin{figure}[htb]
    \centering
    \includegraphics[width=0.4\textwidth]{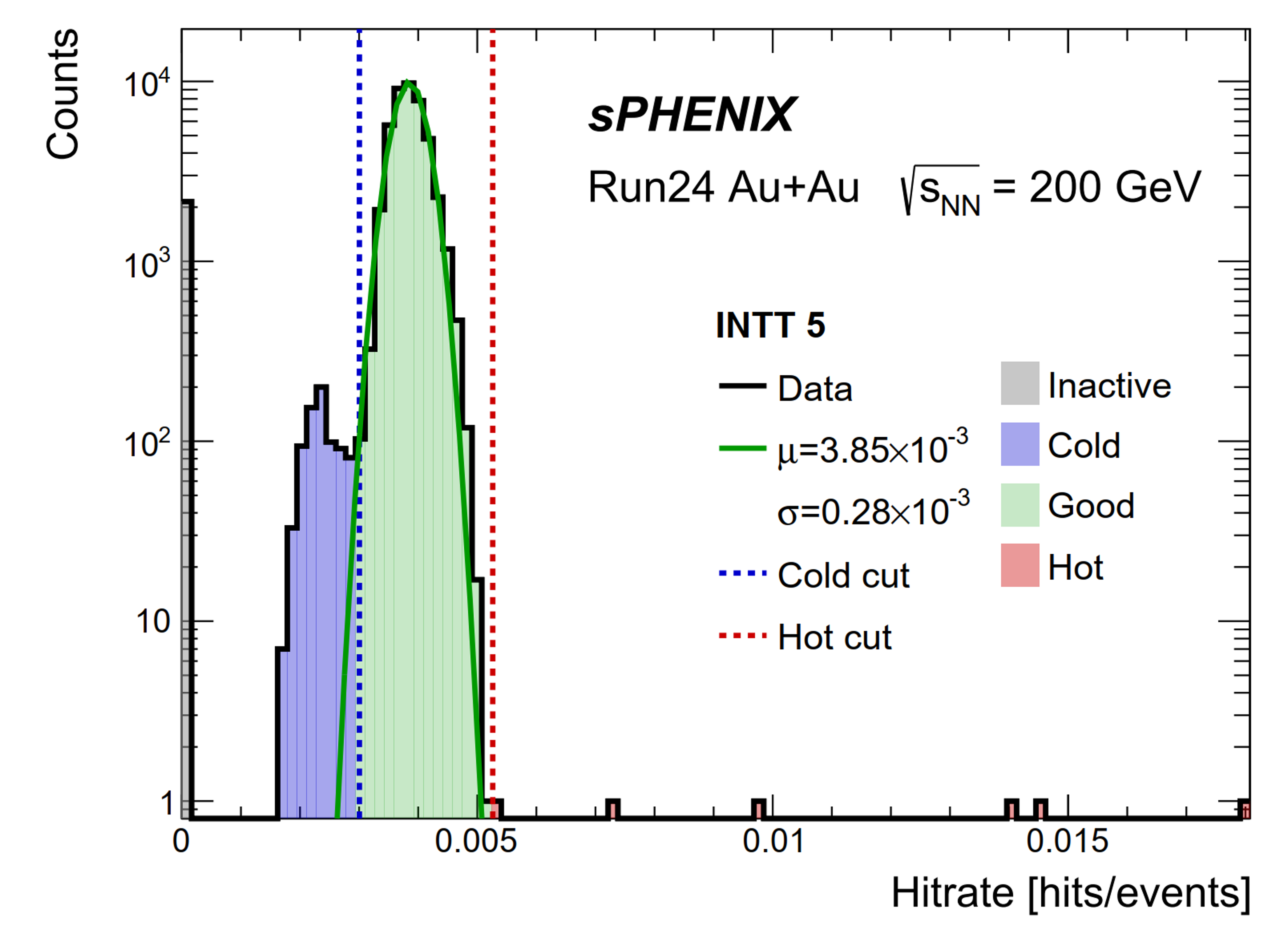}
    \caption{Channel hit-rate distribution for one INTT FELIX server. The solid curve shows the Gaussian fit used to determine the bad-channel classification thresholds. Channels with no hits are classified as inactive, those with hit rates below $\mu - 3\sigma$ as cold, and those with hit rates above $\mu + 5\sigma$ as hot.}
    \label{fig:checkrun_live_channel}
\end{figure}

The live channel fraction of the INTT was evaluated using a data-driven
bad-channel classification based on channel hit-rate distributions. The hit
rate, defined as the number of hits per event, was normalized for geometrical
acceptance effects such as the barrel radius and the strip length of silicon sensor.
For each FELIX server, the corrected hit-rate distribution was fitted with Gaussian functions, and the extracted mean $\mu$ and width $\sigma$ were used to define the classification thresholds.
Figure~\ref{fig:checkrun_live_channel} shows a channel hit-rate distribution for one INTT FELIX server, together with the Gaussian fit and the cold and hot thresholds.
Channels with no hits were classified as inactive, channels with hit
rates below $\mu - 3\sigma$ as cold, and channels with hit rates above
$\mu + 5\sigma$ as hot. All remaining channels were classified as live.
%Channels not assigned to these categories were counted as live channels.

\begin{figure}[htb]
    \centering
    \includegraphics[width=0.48\textwidth]{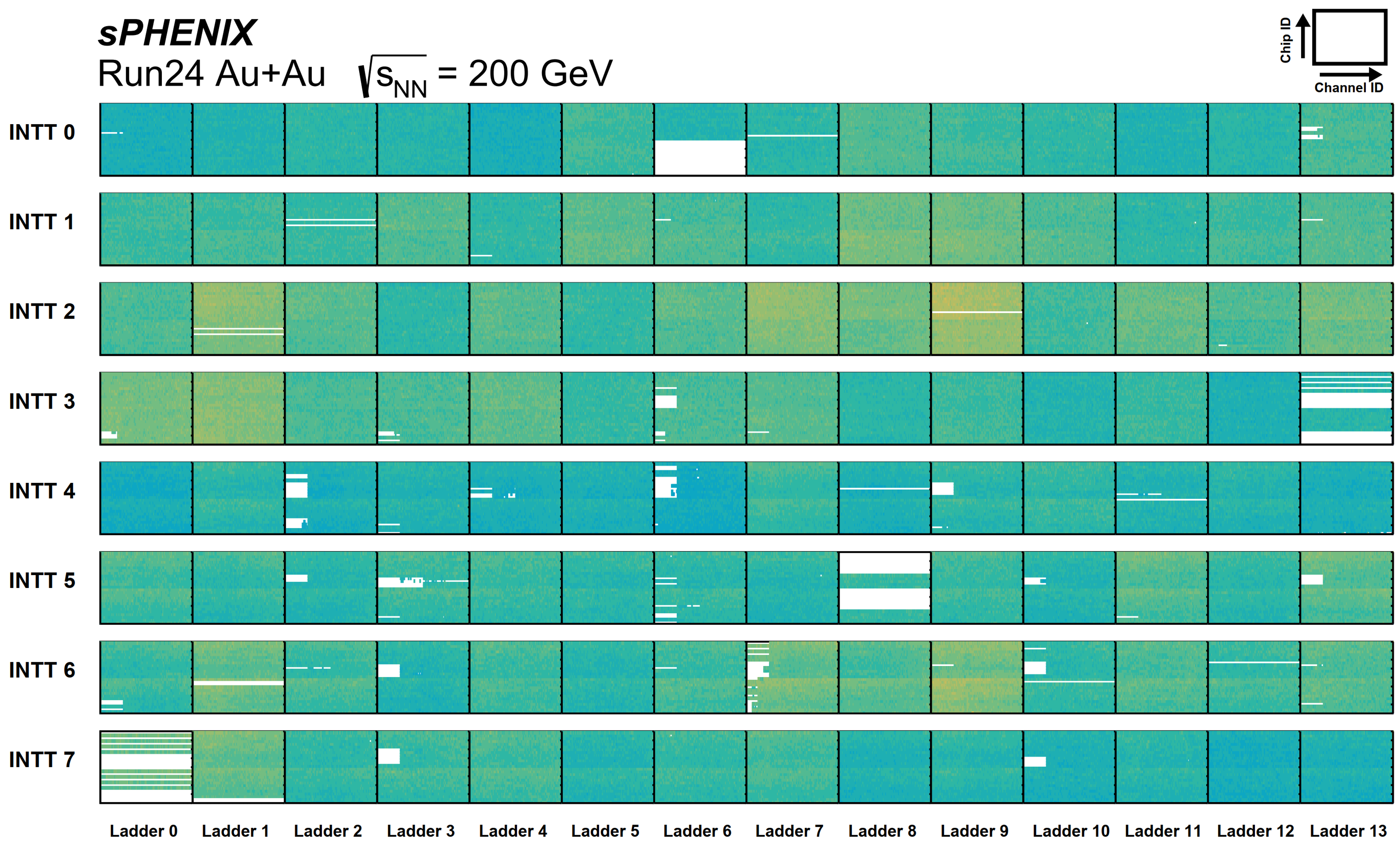}
    \caption{Normalized hit-rate map after applying the bad-channel mask. The map shows the detector response from live channels after inactive, cold, and hot channels are excluded.}
    \label{fig:hitmap_after_mask_live_channel}
\end{figure}

\begin{figure}[htb]
    \centering
    \includegraphics[width=0.45\textwidth]{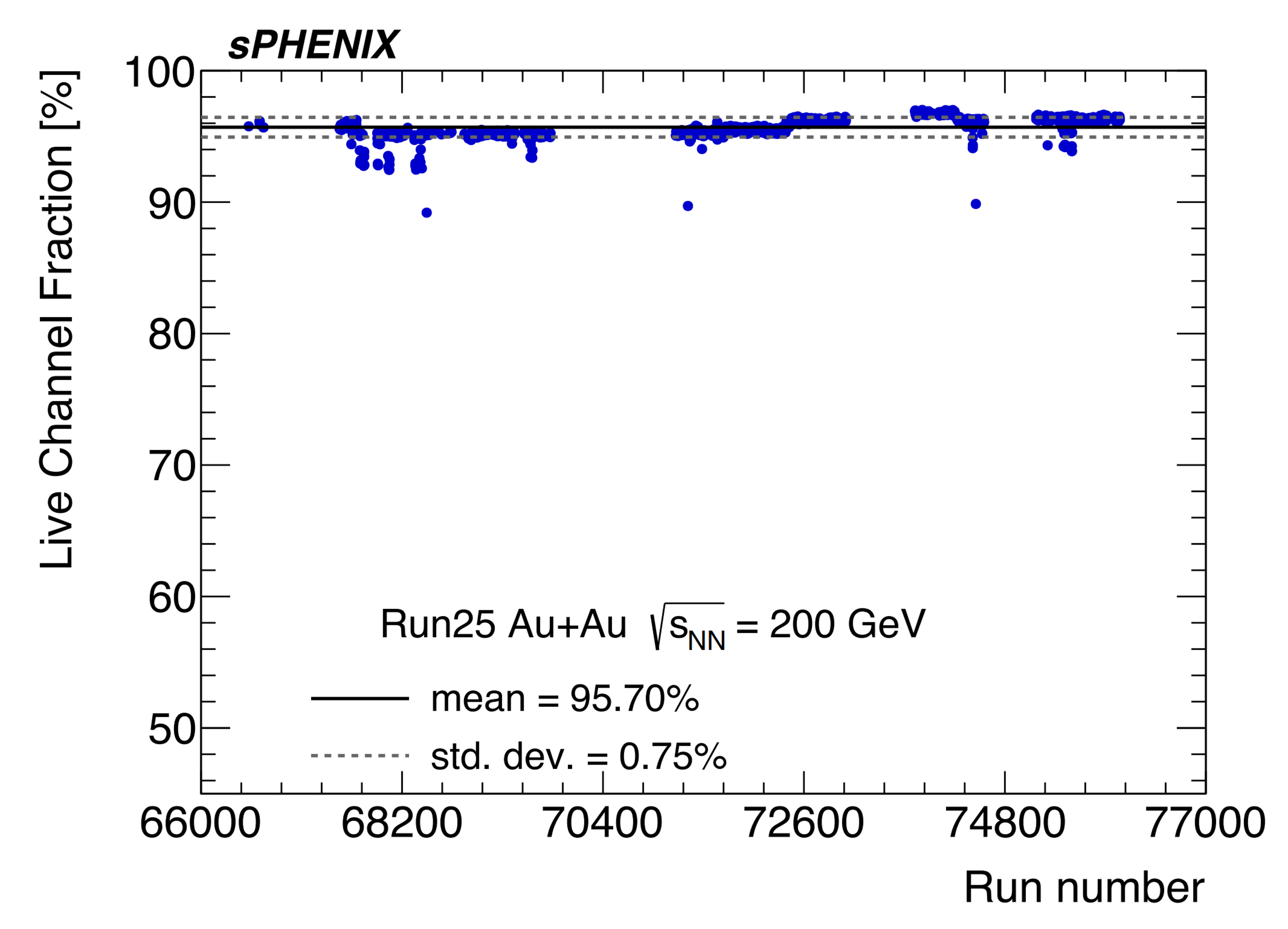}
    \caption{Run-by-run live channel fraction for the Run25 Au+Au. The solid line indicates the average fraction and the dashed lines indicate one standard deviation.}
    \label{fig:good_channel_fraction_run3auau}
\end{figure}

Figure~\ref{fig:hitmap_after_mask_live_channel} shows a channel-by-channel hit-rate map normalized by the number of events after applying the bad-channel mask.
It illustrates that most of the detector remains active after the identified inactive, cold, and hot channels are removed/excluded.
The run-by-run live channel fraction remained stable over the selected Run25 gold-gold (Au+Au) collision data sample, as shown in
Fig.~\ref{fig:good_channel_fraction_run3auau}. The average channel-classification fractions are summarized in Table~\ref{tbl:live_channel_fraction}.

\begin{table}[htb]
\begin{center}
\caption{Average channel-classification fractions for a selected Run25 Au+Au data sample. The quoted standard deviations represent the run-by-run variation.}
\label{tbl:live_channel_fraction}
\begin{tabular}{l|c|c}
\hline\hline
 Category & Mean [\%] & Std. dev. [\%] \\
\hline
 Live     & 95.699 & 0.752 \\
 Inactive &  2.303 & 0.198 \\
 Cold     &  1.815 & 0.699 \\
 Hot      &  0.183 & 0.223 \\
\hline\hline
\end{tabular}
\end{center}
\end{table}

\subsection{INTT Acceptance}
\label{sec_INTT_acceptance}
\begin{figure}
    \centering
    \includegraphics[width=0.45\textwidth]{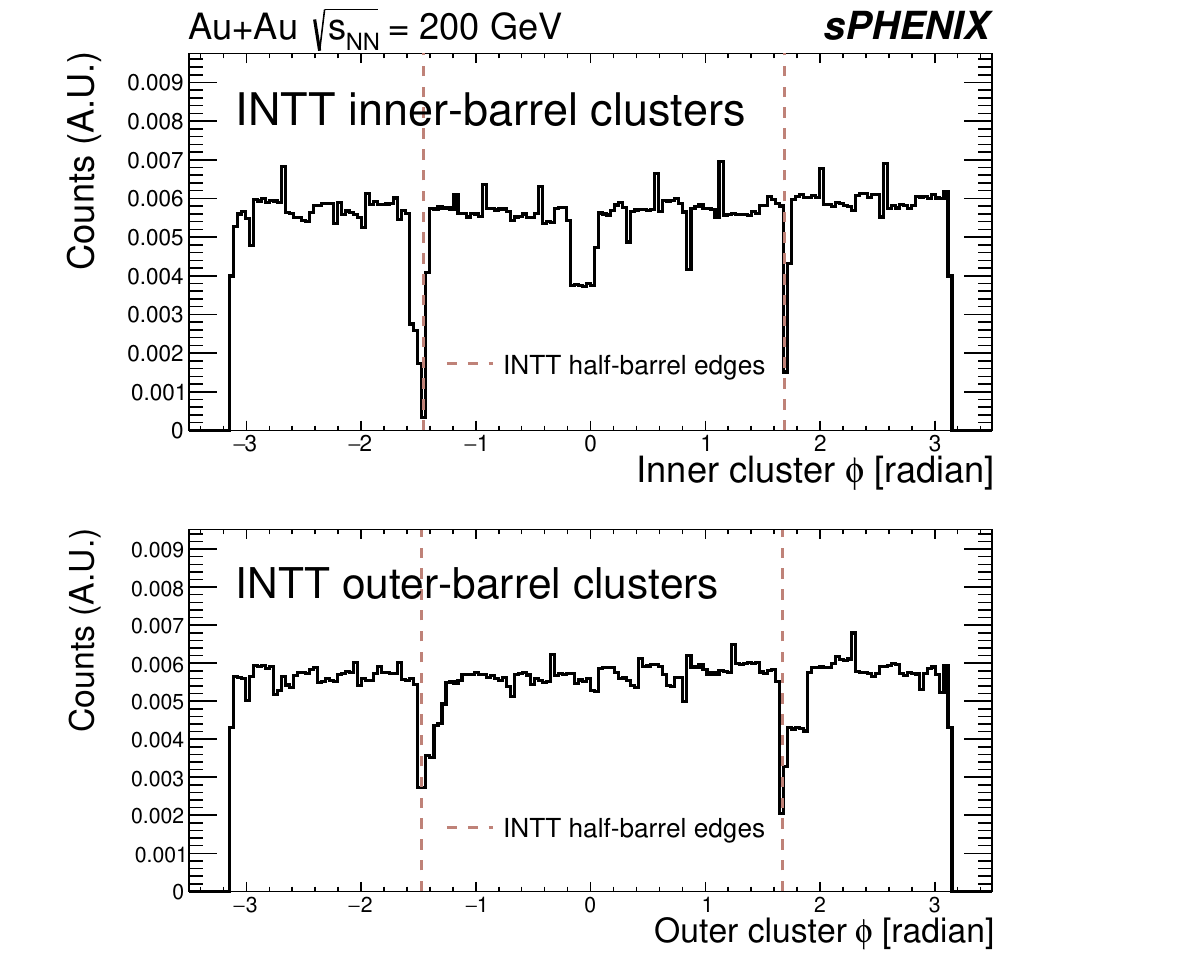}
    \caption{The cluster $\phi$ distribution for the INTT inner (top) and outer (bottom) barrels.}
    \label{fig:cluster_phi}
\end{figure}

Figure~\ref{fig:cluster_phi} shows the cluster $\phi$ distribution for the INTT inner and outer barrels after removing the bad channels identified in Section~\ref{sec_live_bad_chan}.
Besides, hits associated with the triggered crossing are retained. Azimuthal angles of the INTT clusters, formed by grouping adjacent hits in one cell as detailed in~\cite{SHIH2026171312,Abdulhamid2025}, are updated with respect to the determined beam spot, obtained by the method discussed in Section~\ref{beam_spot_reco}. While the distribution is uniform in general, the two large dips observed at cluster $\phi$ approximately -1.5 and 1.8 radians result from the gaps between the two INTT halves, indicating the physical limitation in INTT barrel closure. Other regions with relatively low entries correspond to the azimuthal angles of the malfunctioning sensors, identified by the bad channel classifier.
The periodic peaks arise from the overlap between adjacent ladders as described in Section~\ref{barrel_and_layer}.

\subsection{Hit Multiplicity in Triggered Mode}
\begin{figure} [htb]
    \centering
    \includegraphics[width=0.45\textwidth]{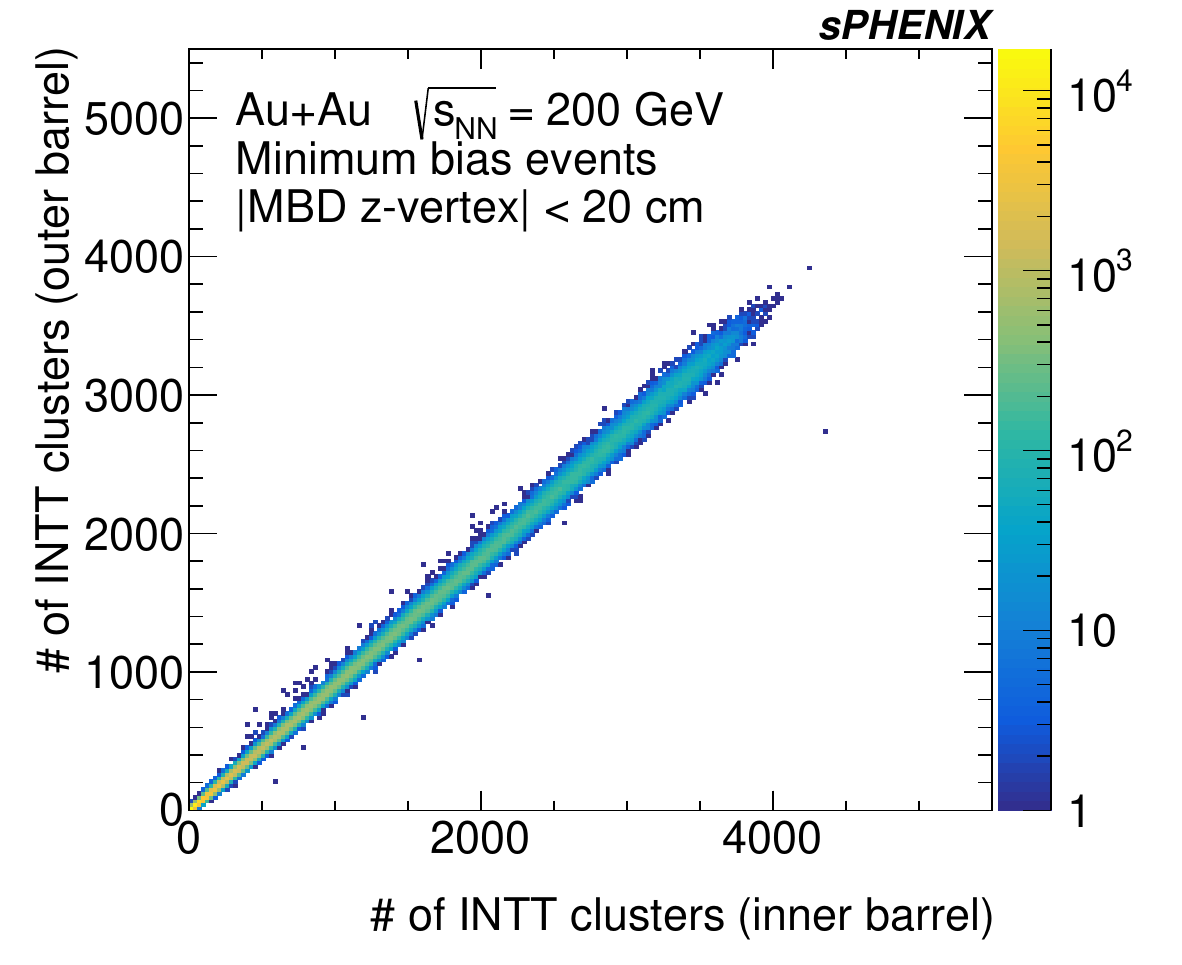}
    \caption{Correlation between the number of INTT clusters in inner and outer barrels in Au+Au collisions at $\sqrt{s_{\mathrm{NN}}}$ = 200 GeV. INTT was operated in triggered mode.}
    \label{fig:corr_INTT_inner_outer_trig}
\end{figure}

\begin{figure} [htb]
    \centering
    \includegraphics[width=0.45\textwidth]{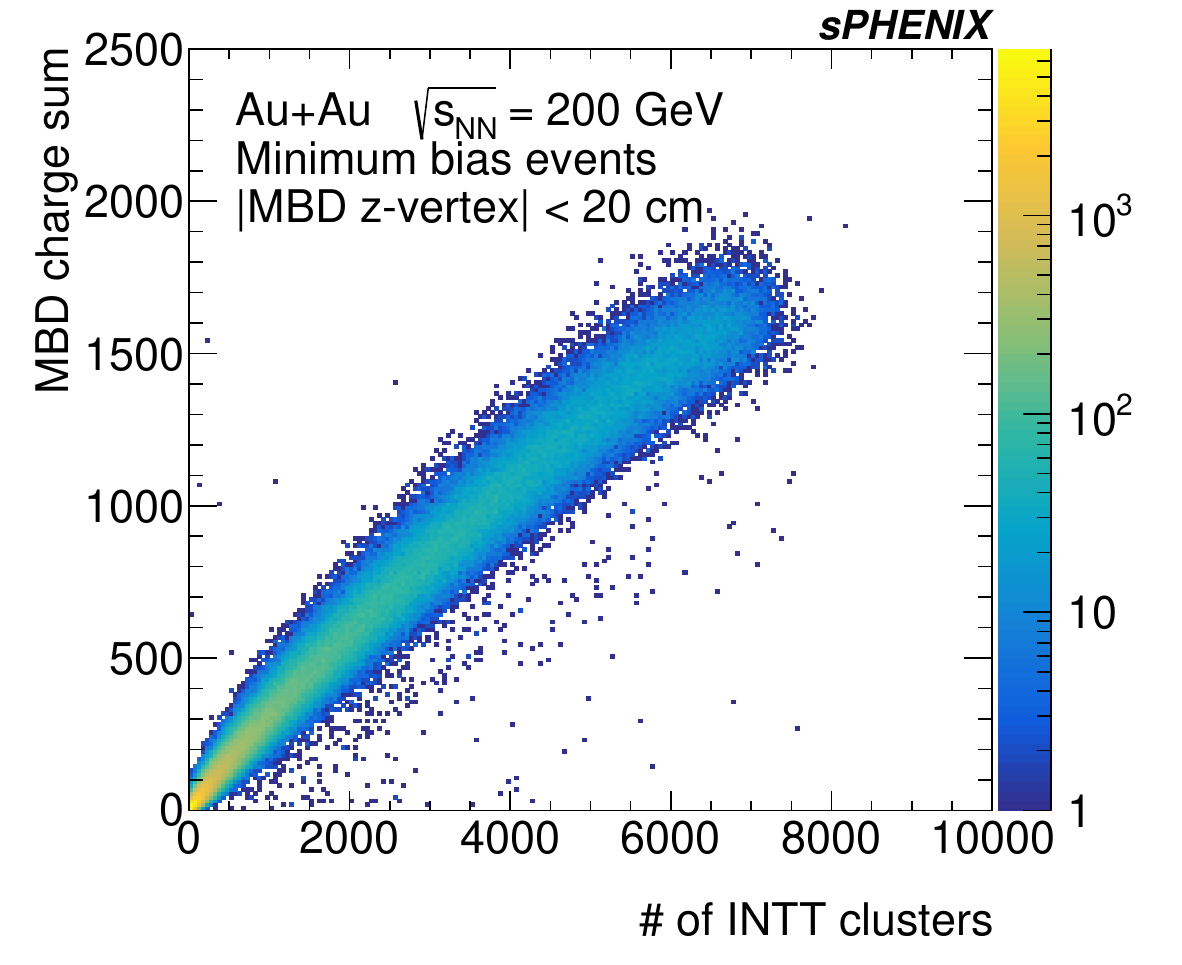}
    \caption{Correlation between the number of INTT total clusters and the MBD charge sum in Au+Au collisions at $\sqrt{s_{\mathrm{NN}}}$ = 200 GeV. INTT was operated in triggered mode.}
    \label{fig:corr_MBD_INTT_trig}
\end{figure}

To benchmark the INTT operation in triggered mode, data collected by INTT and the sPHENIX minimum-bias (MB) detector (MBD) during the RHIC Au+Au collision run are analyzed. MB events with an MBD $z$-vertex within $\pm 20$ cm are retained. Hits associated with the triggered crossing are retained. INTT clusters are counted for the inner and outer barrels, respectively. Their correlation is shown in Fig.~\ref{fig:corr_INTT_inner_outer_trig}. The number of inner clusters exhibits a strong positive correlation with the number of clusters in the outer barrel, validating the synchronization and proper operation of the eight INTT FELIX servers. Figure~\ref{fig:corr_MBD_INTT_trig} shows the correlation between the sum of the numbers of clusters in the INTT inner and outer barrels and MBD charge sum. The MBD is calibrated such that the unity corresponds to one MIP. The observed positive correlation confirms that the INTT signals originate from the same primary interactions as those captured by the MBD, validating the proper synchronization of INTT with other sPHENIX subsystems.
The observed non-linearity is consistent with the expected rapidity dependence of charged-particle production.

\subsection{Streaming-readout Capability}
\begin{figure} [htb]
    \centering
    \includegraphics[width=0.45\textwidth]{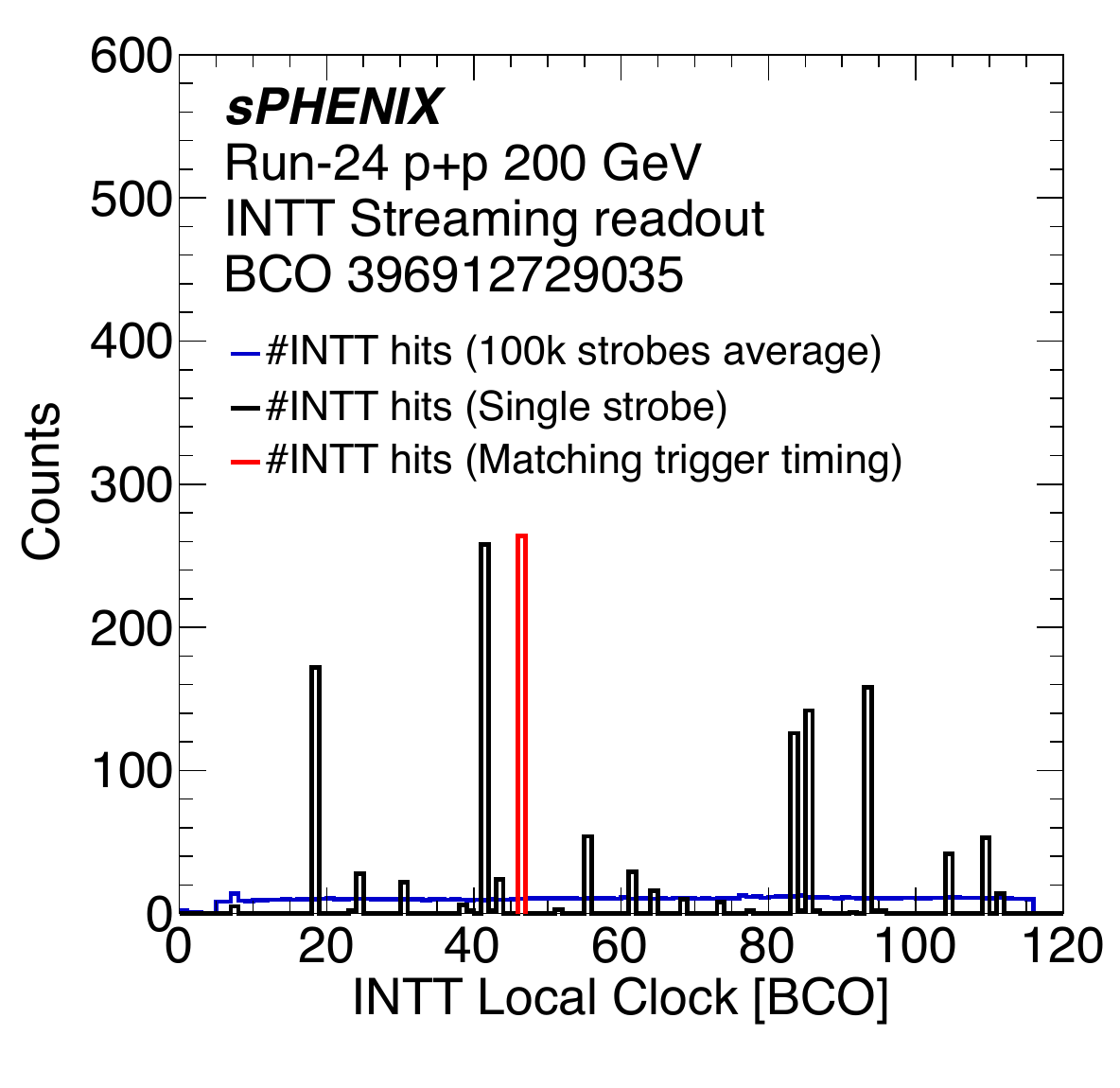}
    \caption{INTT hit counts in the INTT readout frame, referred to as a strobe. The bin in red indicates the collected hits associated with a trigger in a representative strobe. Spikes in black are hits from other collision events in the same strobe. The blue histogram represents the collected hits averaged over 100k strobes.}
    \label{fig:streaming_one_strobe}
\end{figure}

INTT switched to streaming readout mode for data taking in p+p and oxygen-oxygen (O+O) collisions. Figure~\ref{fig:streaming_one_strobe} shows the INTT hit counts in the INTT readout frame, referred to as a strobe. A strobe has the readout period of approximately \SI{13}{\micro\second}, long enough to cover all bunch crossings in one RHIC beam circulation. The red bin in Fig.~\ref{fig:streaming_one_strobe} indicates the collected hits from a bunch crossing associated with a trigger in a representative strobe, while spikes in black represent the recorded hits in the same strobe from other bunch crossings without acquiring a trigger signal, thereby demonstrating the streaming-readout capability of INTT.
The data rate for INTT is $\sim$160\,MB/sec in the streaming mode at the highest p+p collision rate. 

\begin{figure} [htb]
    \centering
    \includegraphics[width=0.45\textwidth]{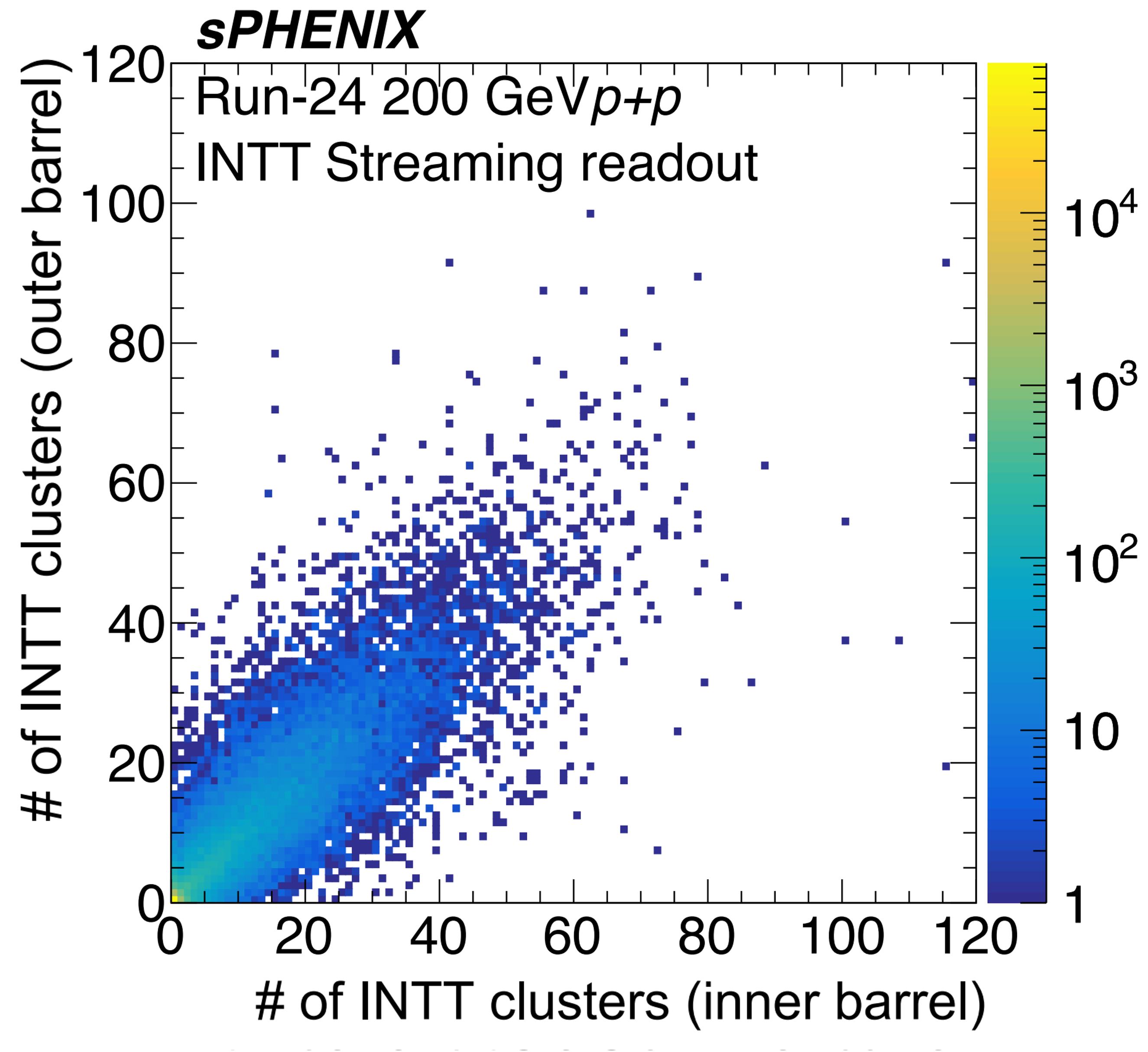}
    \caption{Correlation between the number of INTT clusters in inner and outer barrels in p+p collisions at $\sqrt{s}$ = 200 GeV. INTT was operated in streaming mode.}
    \label{fig:corr_INTT_inner_outer_stream}
\end{figure}

\begin{figure*}[t]
    \centering
    \includegraphics[width=0.9\textwidth]{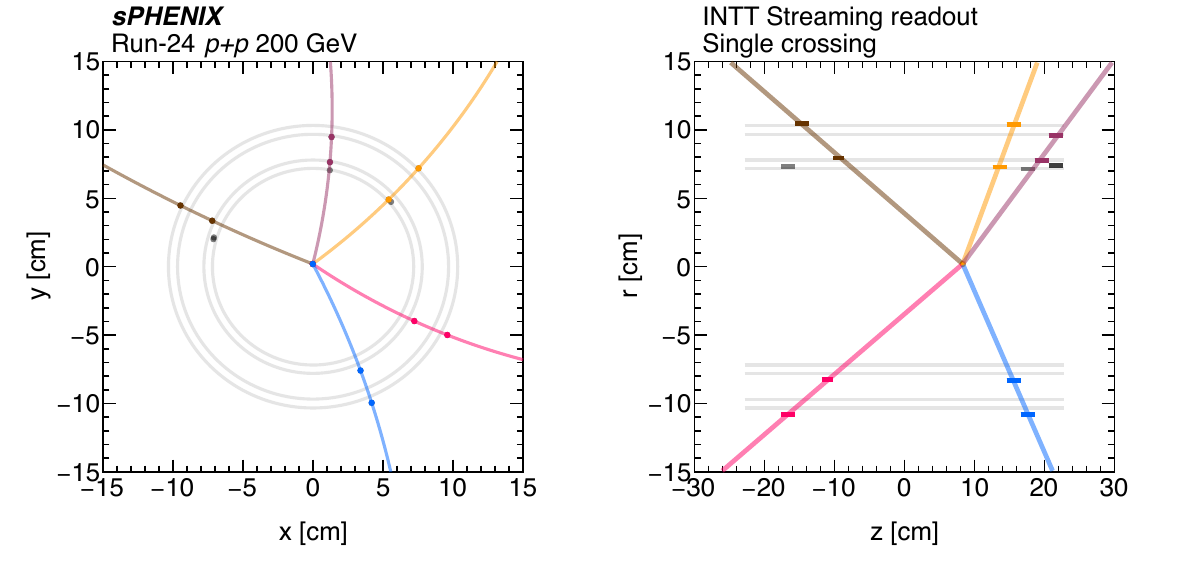}
    \caption{Event display of INTT performing tracking reconstruction using INTT clusters in p+p collisions with INTT in streaming readout mode in (a) the x-y plane and (b) the z-radius plane. 
    The actual hit positions, corrected with the precise survey-based alignment, deviate from the designed positions (gray circles and lines).}
    \label{fig:pp_tracking_eventdisplay}
\end{figure*}

Figure~\ref{fig:corr_INTT_inner_outer_stream} shows the multiplicity correlation between the inner and outer INTT barrels using data collected in p+p collisions at $\sqrt{s}$ = 200 GeV with INTT in streaming readout mode. A clear positive correlation is observed, benchmarking the reliability of its operation in streaming readout mode.
The correlation is less pronounced than that observed in the trigger mode, as shown in Fig.~\ref{fig:corr_INTT_inner_outer_trig}. This difference arises not only from the different collision systems but also from the inclusion of non-triggered events in the streaming readout, such as very low-multiplicity collisions, beam-induced background events, and cosmic rays. These additional event populations dilute the correlation between the inner and outer INTT barrels.
The INTT-standalone tracking analysis is performed to further validate the recorded INTT streaming data. An event display is shown in Fig.~\ref{fig:pp_tracking_eventdisplay}. Particle tracks are reconstructed by utilizing the INTT clusters on top of the vertex-reconstruction capability of INTT, as discussed in Sections~\ref{beam_spot_reco} and~\ref{z_vertex_reco}. The curvatures of charged particles in a magnetic field are clearly identified. The mission of INTT as a tracker is demonstrated.

% \subsection{Trigger Mode Performance} %Ryotaro, Genki
\subsection{Timing} %Ryotaro/Genki
The primary role of the INTT in the sPHENIX tracking system is to assign reconstructed tracks to the correct RHIC bunch crossing.
Accordingly, the detector readout records hit timing only with the BCO granularity (106~ns), and no finer timing information is stored in the data stream.
However, this does not imply that the detector itself is limited to a 1-BCO timing resolution. The intrinsic timing resolution of the INTT is substantially better than 106~ns and can be measured by exploiting the timing characteristics of the FPHX front-end chip.
Specifically, the delay between the INTT local clock and the RHIC beam clock was scanned in steps of (1/6) BCO, thereby shifting the latching time of the digitized hits relative to the beam clock.
By comparing the hit rates at different delay settings, the fine hit-timing distribution was reconstructed with sub-BCO resolution.
This study also determined the optimal latching timing of the FPHX chip, maximizing the fraction of hits assigned to the correct BCO while minimizing the assignment of hits from a single collision to the preceding or subsequent bunch crossings.

The FPHX chip assigns a 7-bit BCO value to each recorded hit.
As the delay setting is varied, hits associated with the same collision appear in different timing bins.
Therefore, a system of linear equations was constructed and solved to reconstruct the fine hit-timing distribution with a bin width of (1/6) BCO.
The linear equations were formulated under the following assumptions:
\begin{enumerate}
\item For a given delay setting, the hit yield measured in each BCO bin is given by the sum of the yields in the corresponding six fine time bins, which were treated as unknown variables.
\item When the delay is changed by one step, the integration range shifts by one fine time bin, corresponding to (1/6) BCO.
\item Hits associated with the triggered bunch crossing are contained within a three-BCO window, with no contribution from outside this window except for a constant offset component.
\end{enumerate}
By combining the equations obtained from the different delay settings, a simultaneous system of linear equations was formed.

\begin{figure}[htb]
    \centering
    \includegraphics[width=0.45\textwidth]{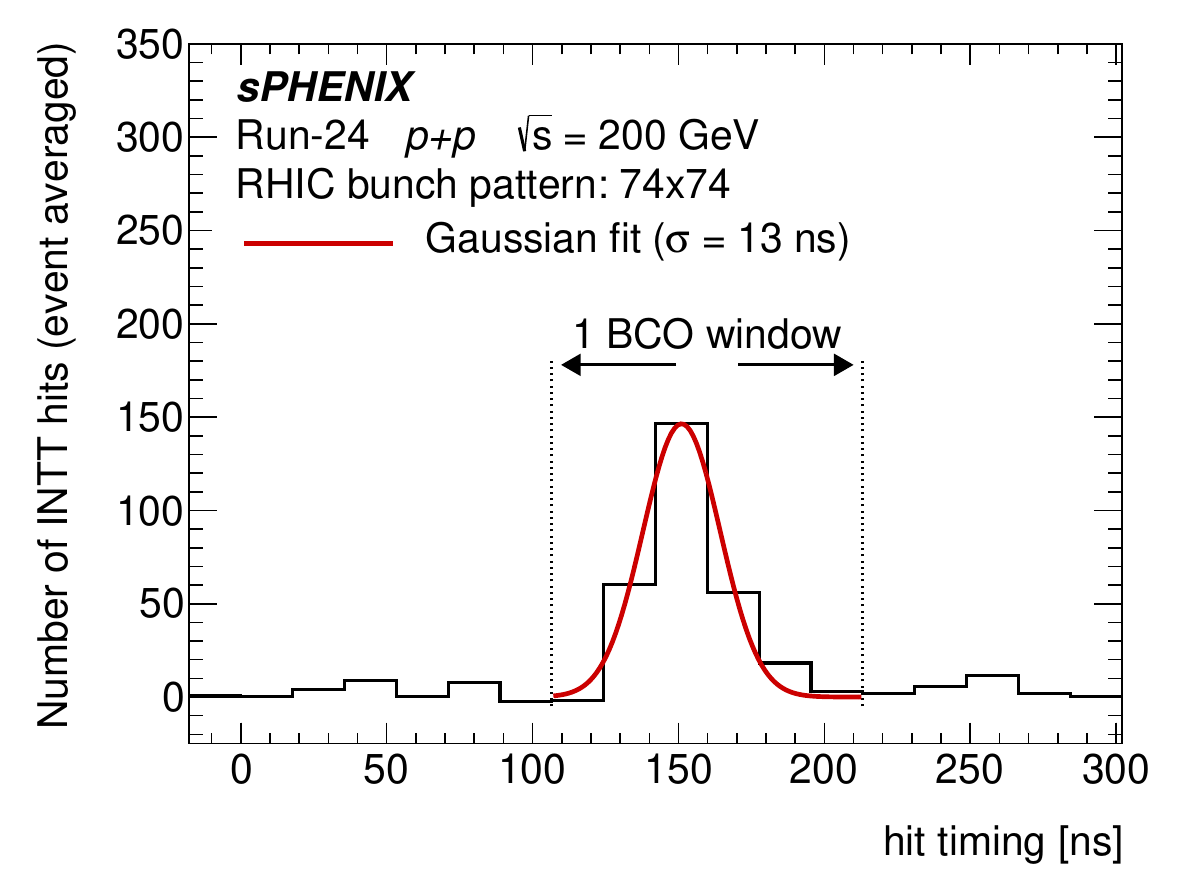}
    \caption{Timing distribution of the INTT hits relative to the RHIC beam clock.}
    \label{fig:TimeResolutionHitTimingDistribution}
\end{figure}

The data used in this analysis were collected during the Run24 p+p commissioning period.
RHIC was operated with the $74\times74$ bunch pattern.
All INTT ladders were included, and only events with high multiplicity, ranging from 200 to 1000 hits, were used.
The extracted distribution is shown in Fig.~\ref{fig:TimeResolutionHitTimingDistribution}.
A clear peak corresponding to hits from the triggered bunch crossing is observed.
The 1-BCO timing window used during the Run24 physics data-taking period is indicated by the dotted lines.
A Gaussian fit was performed to characterize the distribution, and the standard deviation of $13.24 \pm 0.001~(\text{stat.}) \pm 1.89~(\text{sys.})$~ns was obtained.
The intrinsic time resolution is confirmed to be good enough for the crossing determination, and most hits are contained within the 1-BCO window.

\subsection{Noise Fraction} % Cheng-Wei/Genki
The noise contamination in INTT that affects the offline analysis is evaluated using a data-driven approach based on the INTT data collected in streaming readout mode with p+p collisions. During the data taking, the nominal threshold was set, and the RHIC bunch configuration consisted of 12 crossings filled with beam particles. The 12 filled crossings are each separated by nine empty crossings, resulting in a bunch crossing interval of 10 BCOs. The streaming mode of INTT opens a readout window long enough to cover all crossings in a RHIC circulation and collects all analog signals that surpass the threshold without requiring a trigger signal. Therefore, hits collected at crossings without particle bunches enable an assessment of the INTT noise level, as those hits are originated from electronic noise or lingering slow and/or thermal neutron background within the detector that is not associated with collision events.  

\begin{figure} [htb]
    \centering
    \includegraphics[width=0.45\textwidth]{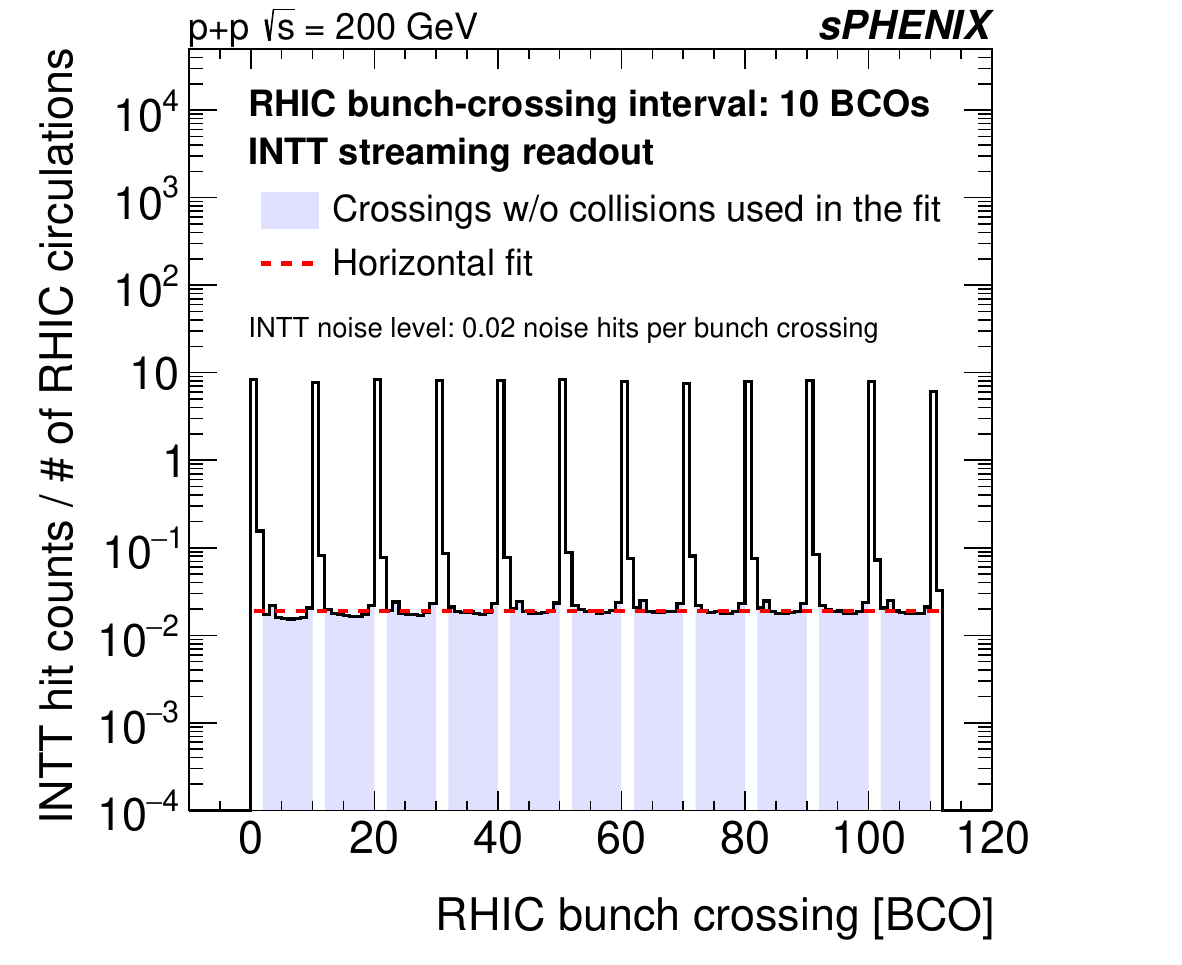}
    \caption{Number of INTT hits, normalized by the number of RHIC circulations, for each crossing in the RHIC circulation.}
    \label{fig:h1D_NoiseFraction}
\end{figure}

Events that meet the MB trigger requirement with MBD z-vertex within $\pm$13.3\,cm and with at least one INTT hit associated with the triggered crossing are analyzed. The analysis focuses on the noise fraction of channels used for future physics analyses. Therefore, bad channels, identified by the method described in Section~\ref{sec_live_bad_chan}, are removed. Figure~\ref{fig:h1D_NoiseFraction} shows the number of INTT hits collected in streaming mode, normalized by the number of RHIC circulations, for each crossing in the RHIC circulation. The ten spikes in Fig.~\ref{fig:h1D_NoiseFraction} correspond to the ten RHIC crossings with beam particles. The shoulder entries next to the spikes consist of signal hits migrated from the spikes in addition to the noise hits. The fraction of signal-hit migration is typically below 1\,\%, which is expected from the intrinsic timing resolution of the INTT and optimized delay timing.
A horizontal line is fit to the regions without hits associated with collisions, denoted as bins filled with light purple in Fig.~\ref{fig:h1D_NoiseFraction}. The level of the horizontal line, yielding 0.02, represents the number of noise hits in INTT per bunch crossing.

\subsection{INTT Hit Detection Efficiency}
\begin{figure} [htb]
    \centering
    \includegraphics[width=0.45\textwidth]{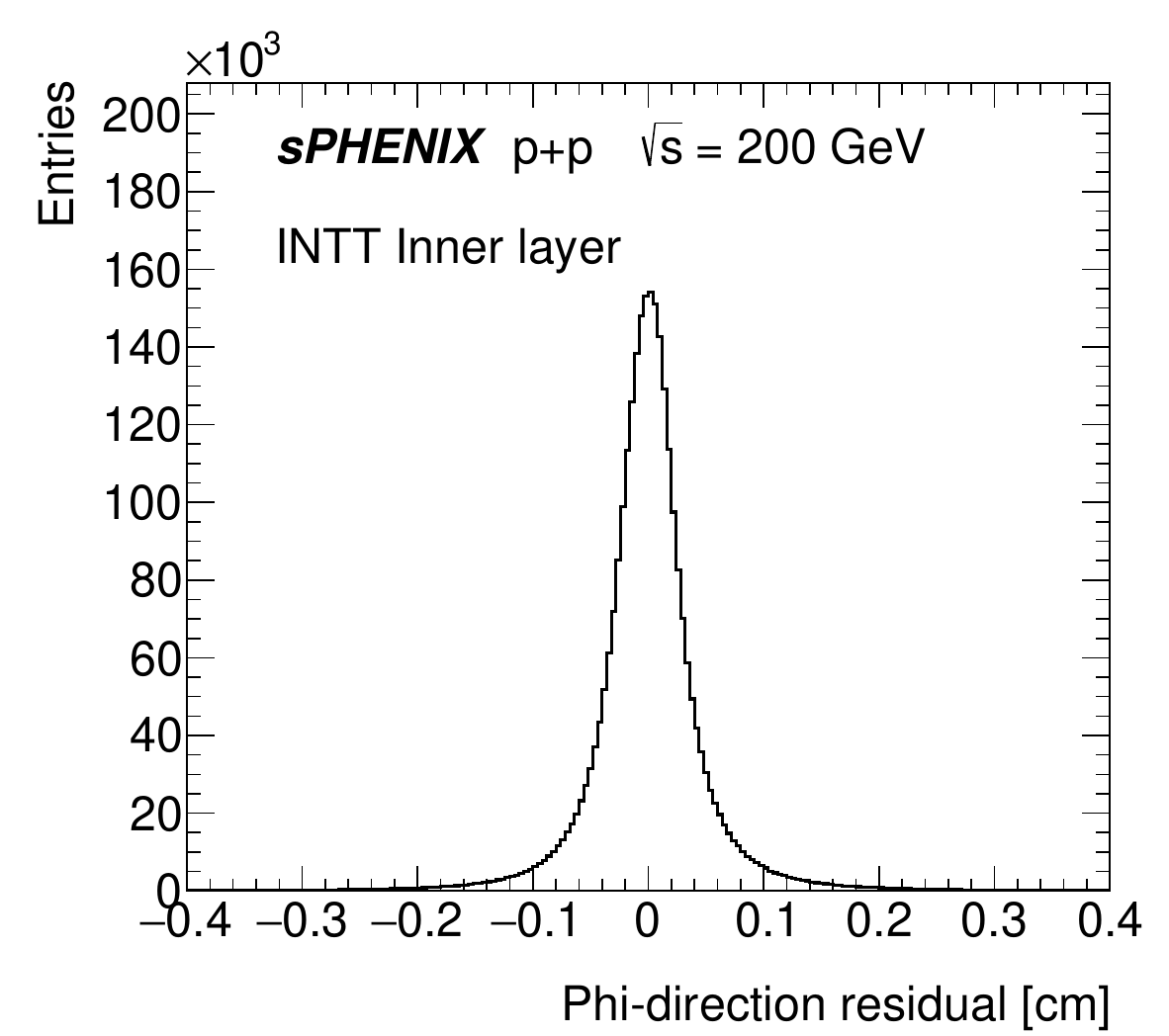}
    \caption{Phi direction residual distribution of one representative INTT half-ladder.}
    \label{fig:residual}
\end{figure}

The INTT provides hit information with an efficiency close to 100\,\% for charged particles traversing the active sensor area. A hit detection efficiency exceeding 99.5\,\% was demonstrated using a pre-production INTT ladder in a beam test performed in 2021~\cite{SHIH2026171312}. To verify that this performance was maintained after detector construction, the hit detection efficiency was re-evaluated using p+p collision data taken with zero magnetic field after the barrel assembly. The efficiency of the inner INTT layer was measured using tracks reconstructed from the three MVTX layers and the outer INTT layer, ensuring that charged particles traversed the target sensor. Figure~\ref{fig:residual} shows the residual distribution in the channel-column direction for a representative INTT half-ladder. A hit detection efficiency of $(99.8 \pm 0.03)\%$ was obtained within a residual window of $\pm0.1$~cm for clusters that passed fiducial cuts away from the chip edges, confirming that the excellent performance demonstrated with the pre-production ladder was successfully reproduced in the fully assembled detector.

\subsection{Cosmic-ray Event Display}
\begin{figure} [htb]
    \centering
    \includegraphics[width=0.45\textwidth]{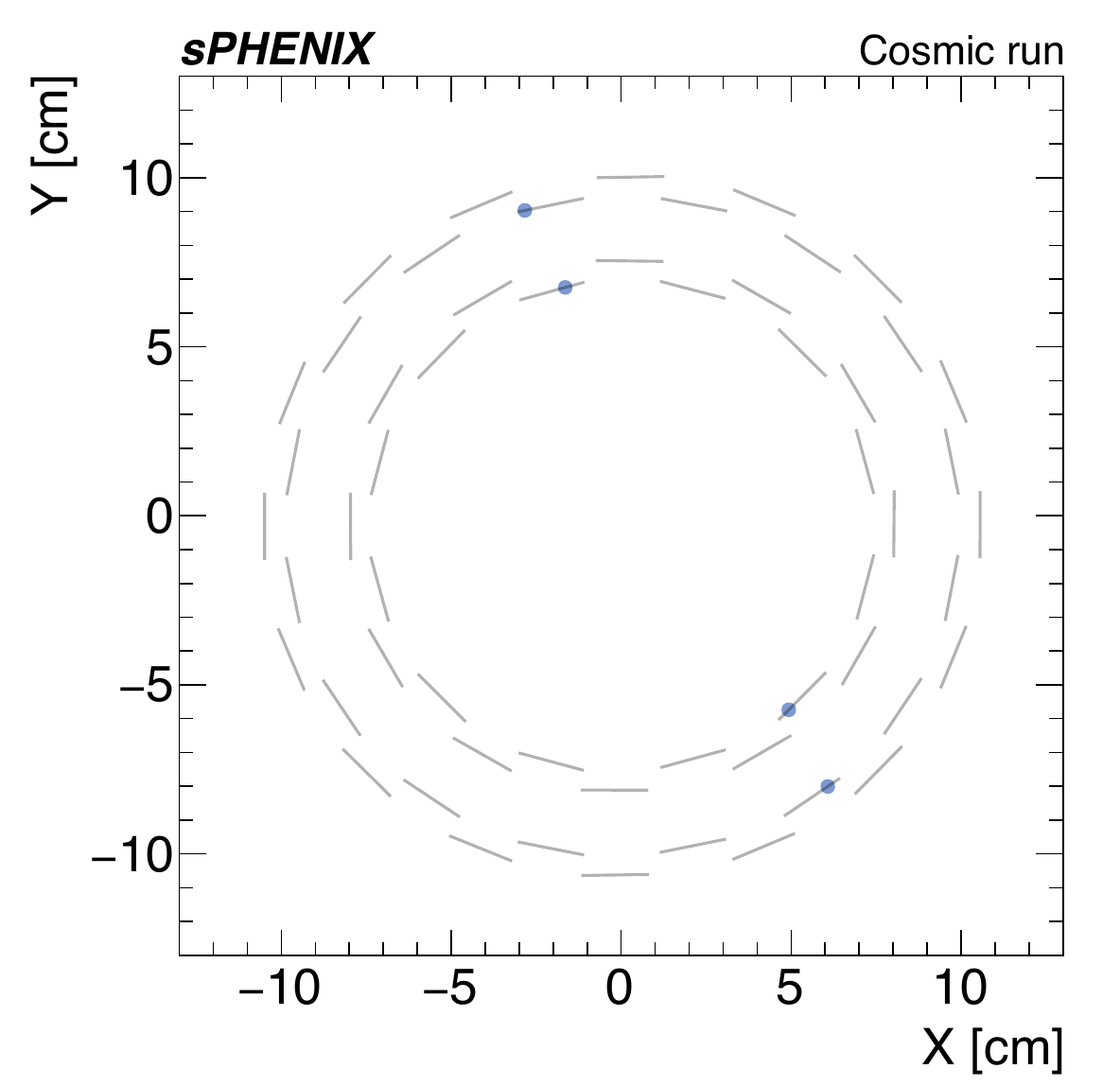} \\
    \includegraphics[width=0.45\textwidth]{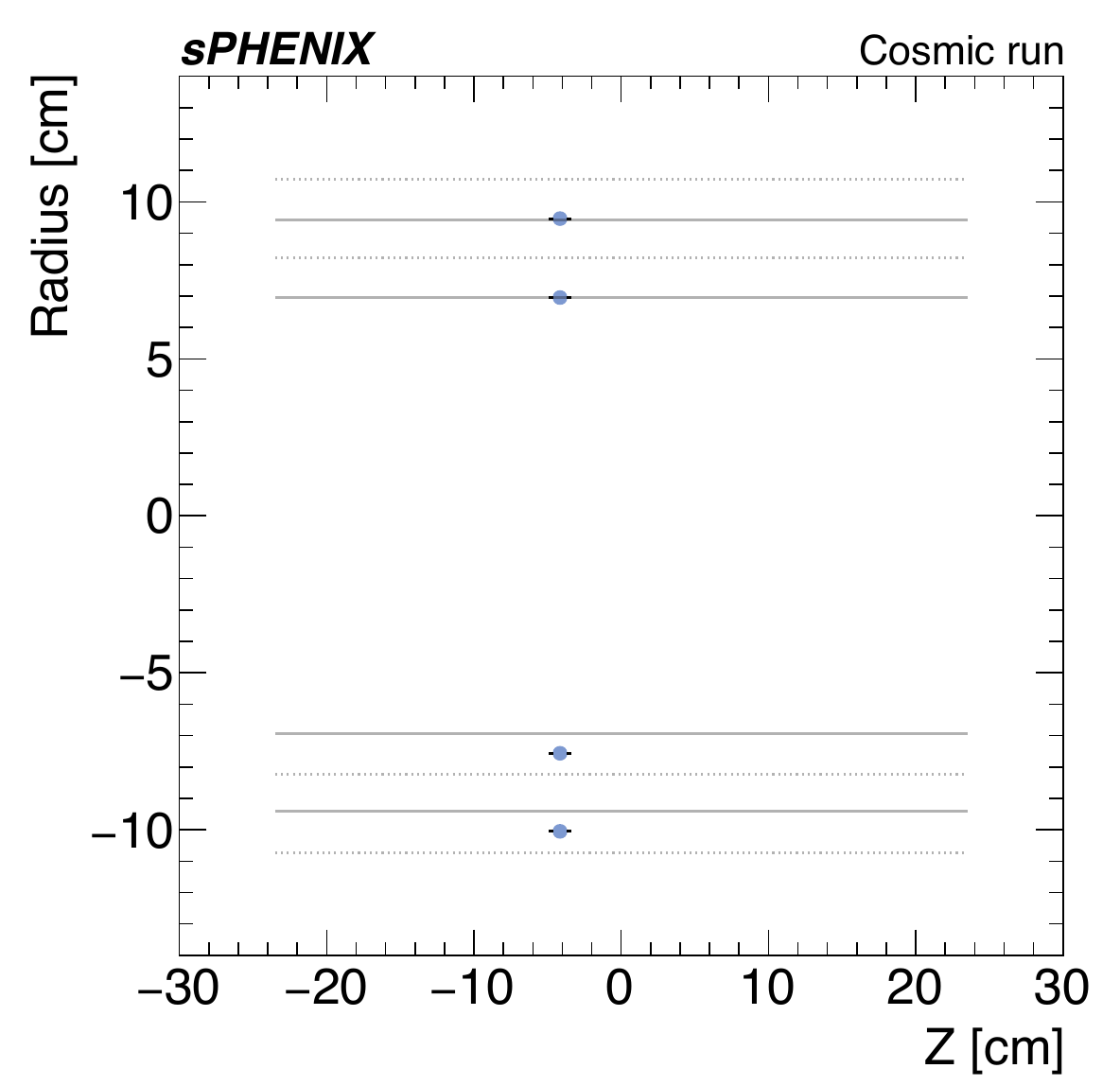}
    \caption{Event display of one cosmic-track event observed by INTT in the x-y (top) and z-radius (bottom) planes. The negative radius represents clusters with the azimuthal angles $\phi=[-\pi,0]$. The horizontal error bars in the bottom panel indicate the stripe lengths of INTT.}
    \label{fig:cosmic_evt_display_trig}
\end{figure}

Cosmic data collected during the 2023 sPHENIX commissioning period are analyzed with the goal of identifying clear cosmic-ray tracks, providing another important benchmark to confirm the integrity of INTT recorded data at the beginning of the sPHENIX operation. Data with INTT in triggered mode are used.  The trigger signal is provided by the coincidence of the center region of the top and bottom of oHCal. The cosmic-ray search focuses on single-track events. Given the ladder arrangement of INTT, cosmic rays typically traverse at least four ladders in most cases.  Therefore, events with more than seven clusters or fewer than four clusters are discarded. Cosmic tracking is performed by preparing all possible cluster combinations with 4 to 7 clusters. All the combinations are evaluated by fitting the clusters with a straight line in the x-y and z-radius planes. The most linear combination with a reduced $\chi^{2}$ smaller than 5 in both planes is considered a cosmic track. Figure~\ref{fig:cosmic_evt_display_trig} shows an example of a cosmic event display in the x-y and z-radius planes. The successful reconstruction of cosmic-ray tracks provides an important commissioning benchmark, confirming both the integrity of the recorded data and the capability of the INTT to perform precision charged-particle tracking in the sPHENIX experiment.

\subsection{INTT-standalone beam spot reconstruction} %Cheng-Wei
\label{beam_spot_reco}
Detailed analyses are performed utilizing the correlation between the inner- and outer-barrel clusters to better understand the collected INTT data. The reconstruction of the beam spot, defined as the vertex position in the transverse plane averaged across multiple events, is developed using the INTT-standalone data taken in Au+Au collisions with zero magnetic field and is described as follows. 

\begin{figure} [htb]
    \centering
    \includegraphics[width=0.48\textwidth]{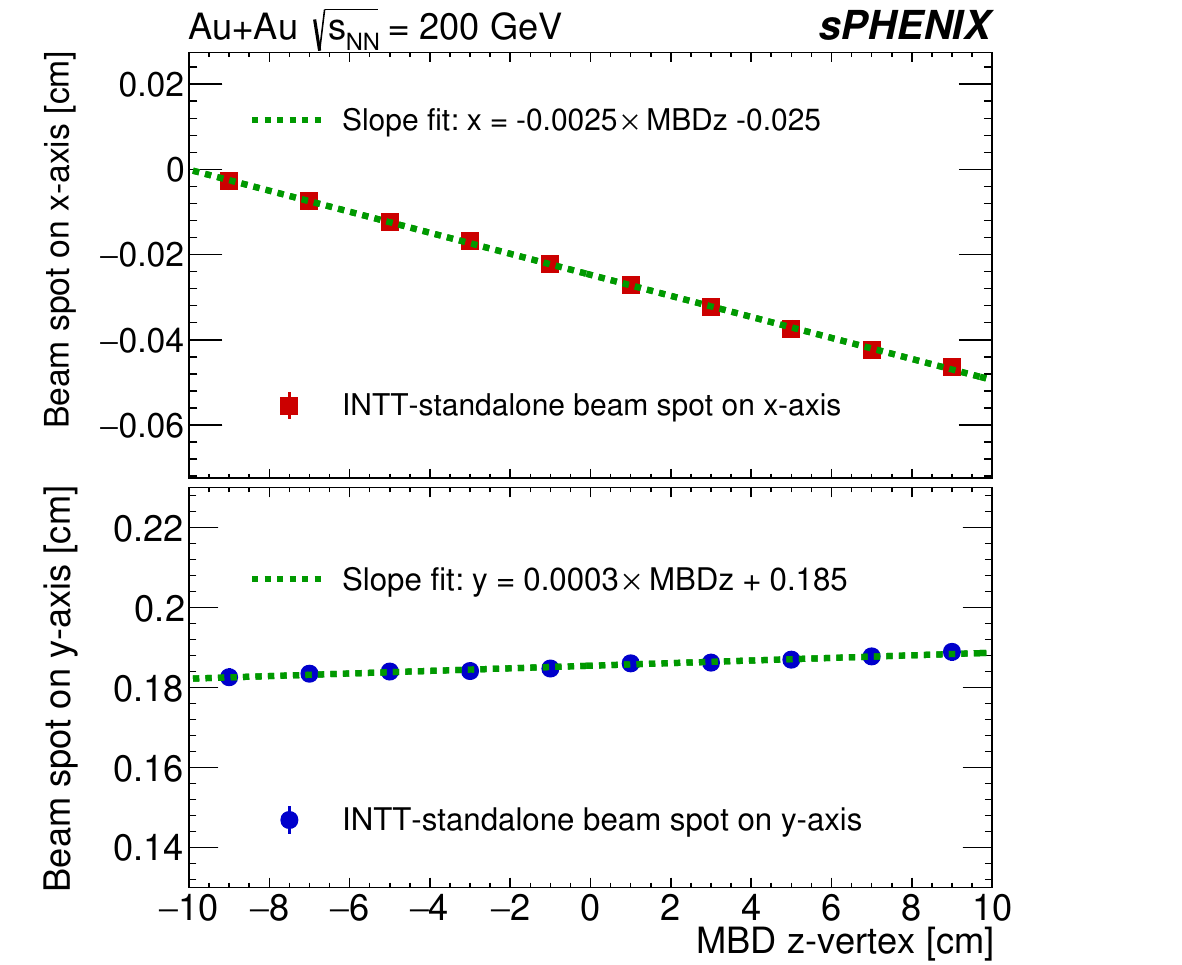}
    \caption{INTT-standalone event-averaged beam spot on the x-axis (top panel) and the y-axis (bottom panel) as a function of MBD z-vertex.}
    \label{fig:beamspot_MBDz}
\end{figure}

A finely segmented two-dimensional histogram, with dimensions large enough to contain the truth beam spot position, is prepared. MB events with a z-vertex reconstructed by MBD within $\pm10$~cm are retained. The inner-barrel clusters are paired with all outer-barrel clusters. Cluster pairs having an azimuthal angular difference within $\pm45$ degrees are retained, and their trajectories are populated into the histogram. The histogram therefore consists of the populations of signal cluster pairs associated with the charged particles emitted from collisions, as well as contributions from combinatorial cluster pairs. The combinatorial background is removed by discarding bins with a content below 70\% of the maximum bin content. The beam spot position is obtained as the mean values of the residual bins on both the x- and y-axes. In the analysis, the beam spot is determined as a function of the MBD z-vertex for every 2 cm, as shown in Fig.~\ref{fig:beamspot_MBDz}.
A clear $z$-dependent beam spot displacement along the $x$-axis is observed. The measured dependence is in good agreement with the expected beam trajectory for the intentional 2~mrad RHIC beam crossing angle, demonstrating the validity of the INTT-based beam spot reconstruction.

\subsection{INTT-standalone Z-vertex Reconstruction} % Cheng-Wei
\label{z_vertex_reco}
The INTT, a barrel silicon strip tracker, was not primarily designed for primary vertex reconstruction along the z-axis, as its strip pitch in the $z$ direction is relatively coarse (16\,mm and 20\,mm)\cite{Nakagawa2025}. Nevertheless, its excellent spatial resolution and large geometrical acceptance enable reconstruction of the primary collision vertex using the INTT alone.
%INTT, the barrel strip tracker, is not conventionally designed for reconstructing the primary collision vertex.
Its potential is investigated using INTT-standalone data taken in Au+Au collisions with zero magnetic field, since it can provide a valuable reference for calibrating the z-vertex measured by MBD. The algorithm is briefly described as follows. 

\begin{figure} [htb]
    \centering
    \includegraphics[width=0.45\textwidth]{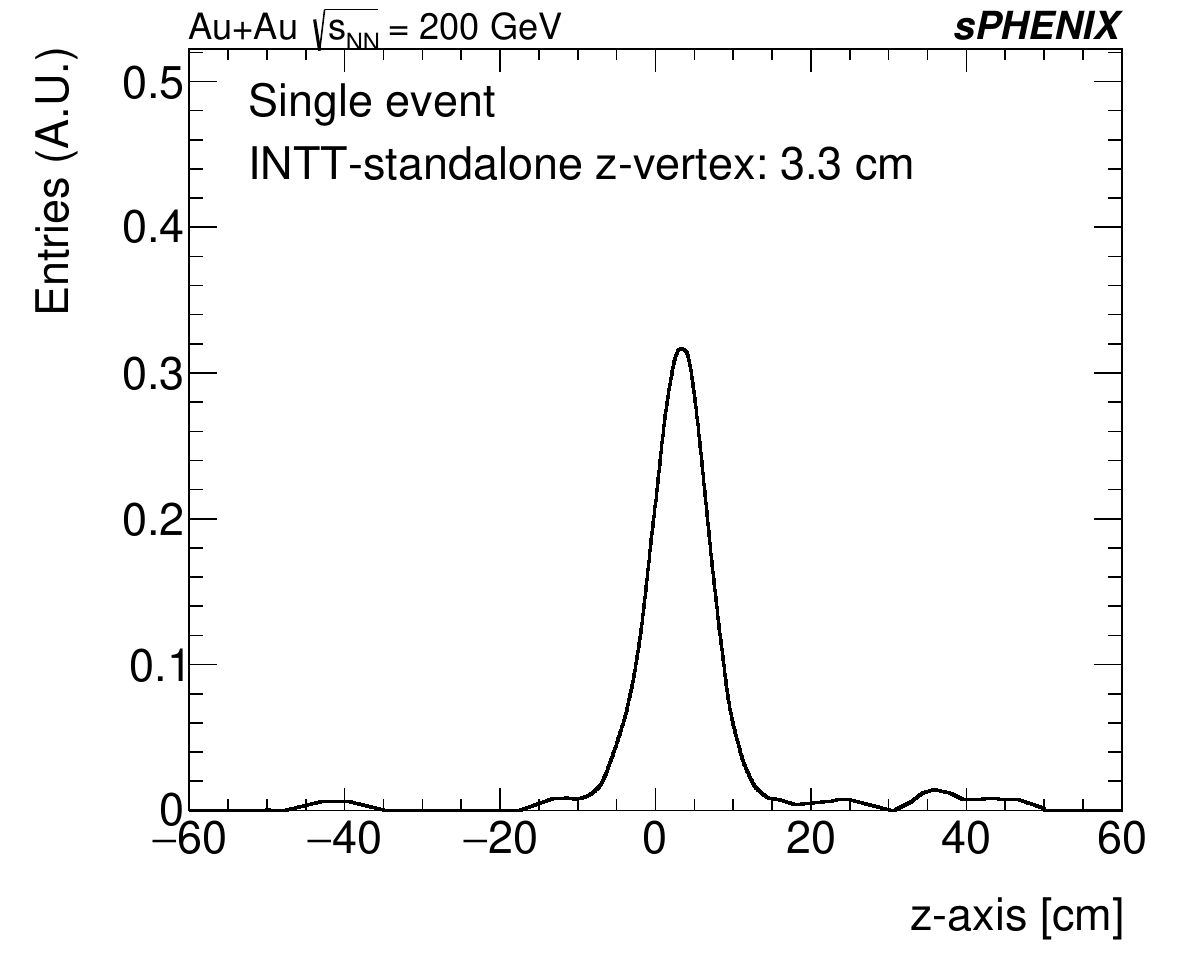}
    \caption{The Gaussian-like distribution formed by stacking up the trapezoids of the valid cluster pairs in a single Au+Au collision event.}
    \label{fig:zvtx_evt_display}
\end{figure}

In one event, cluster pairs with a small azimuthal angular difference calculated with respect to the beam spot, averaged over the selected MBD z-vertex range, are retained and referred to as the valid cluster pair. Since a particle can interact with an INTT strip at any position along its length along the z-axis, the probability distribution of collision location inferred from a valid cluster pair exhibits the shape of a trapezoid. The trapezoids of the valid cluster pairs in a single event are stacked, forming a Gaussian-like distribution as shown in Fig.~\ref{fig:zvtx_evt_display}. The peak of the Gaussian-like distribution is used to determine the event primary z-vertex.

\begin{figure} [htb]
    \centering
    \includegraphics[width=0.48\textwidth]{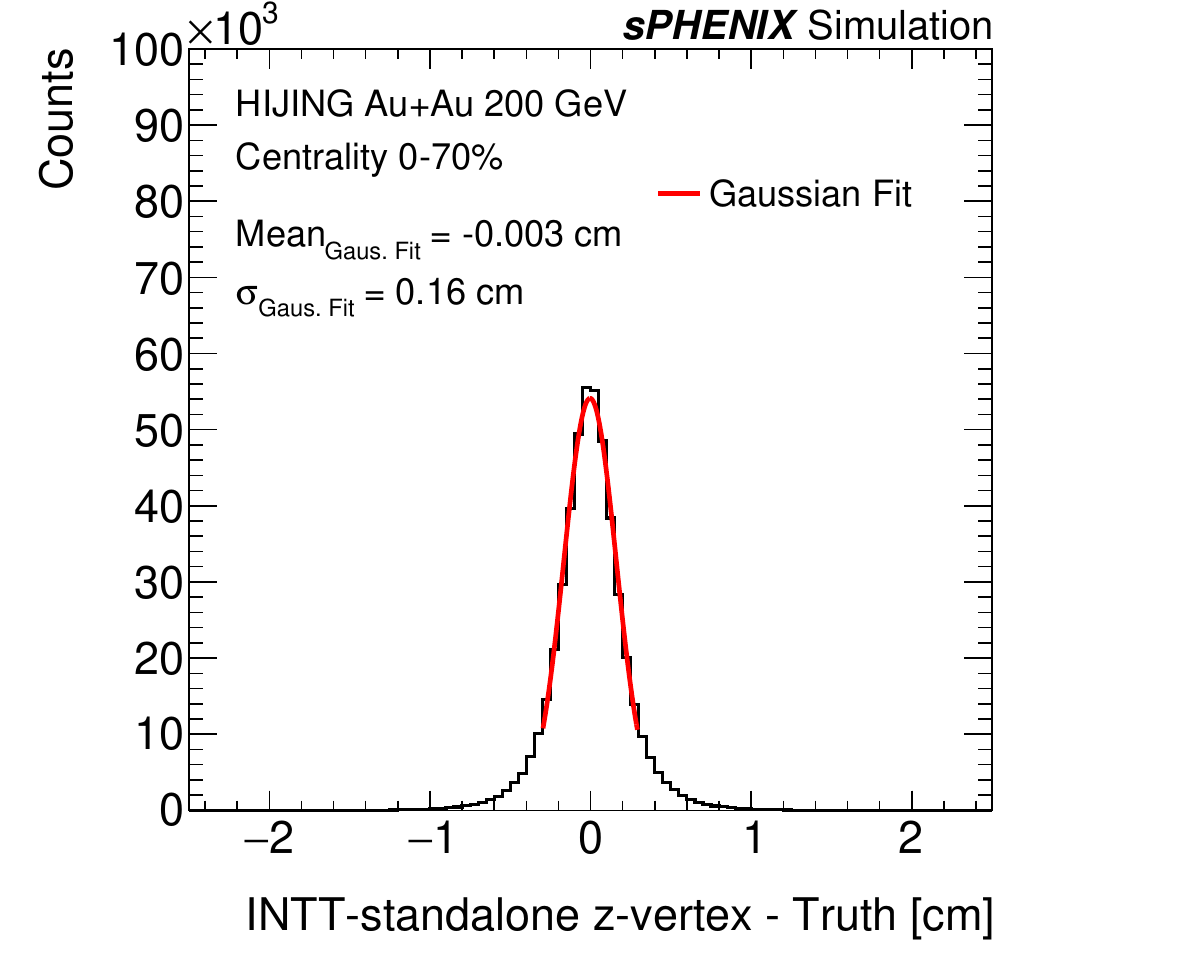}
    \caption{Distribution of the difference between the INTT-standalone reconstructed z-vertex and the truth with HIJING simulation events.}
    \label{fig:zvtx_resolution_sim}
\end{figure}

\begin{figure} [htb]
    \centering
    \includegraphics[width=0.48\textwidth]{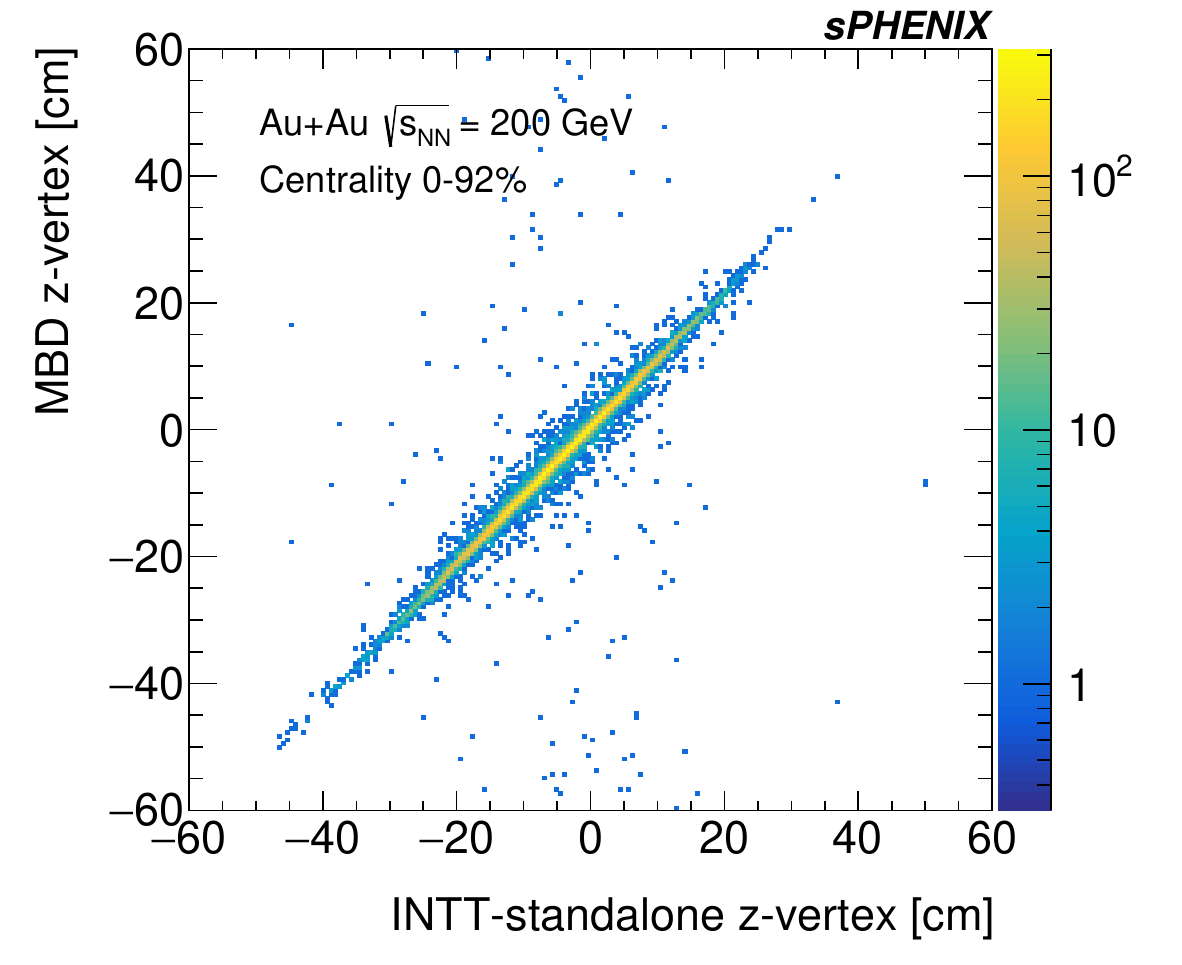}
    \caption{Correlation between the z-vertices reconstructed by INTT and MBD in Au+Au collisions.}
    \label{fig:zvtx_correlation_data}
\end{figure}

The performance of this reconstruction approach is evaluated with HIJING simulation samples, as shown in Fig.~\ref{fig:zvtx_resolution_sim}. The reconstruction resolution can reach up to 1.6 mm in events with centrality class 0-70\,\%, which is 10 times smaller than the intrinsic strip length of INTT. The method is further applied to data and compared with the z-vertex reconstructed by MBD, as shown in Fig.~\ref{fig:zvtx_correlation_data}. The result exhibits a positive correlation over $\pm50$ cm in the z-axis, offering an additional confirmation of the integrity of the INTT data. The beam spot determination and the per-event z-vertex reconstruction described in this paper serve as the foundation of the physics measurement in~\cite{Abdulhamid2025}.

%%%%%%%%%%%%%%%%%%%%%%%%%%%%%%%%%%%%%%%%%%%%%%%%%%
%%%%%%%%%%%%%%%% Summary %%%%%%%%%%%%%%%%%%%%%%%
%%%%%%%%%%%%%%%%%%%%%%%%%%%%%%%%%%%%%%%%%%%%%%%%%%
\section{Summary}
In summary, we present the design, construction and commissioning performance of the INTT, a barrel-type silicon-strip tracker built for the sPHENIX experiment at RHIC. The INTT barrel occupies the middle layer of the sPHENIX tracking system between the MVTX and the TPC, and consists of two concentric cylindrical barrels (24 inner and 32 outer silicon ladders) with low-mass carbon-fiber support structure, providing full azimuthal coverage over $|\eta|<1.07$ for vertices within ±10 cm of the interaction point. The detector's data acquisition is built by employing FELIX boards and servers, supporting both triggered and streaming readout modes, with precise clock distribution and synchronization to RHIC's clock.

Performance results from Run24 and Run25 Au+Au, p+p, and O+O collision data demonstrate stable and reliable operation: an average live-channel fraction of $\sim95.7\,\%$, uniform azimuthal cluster acceptance, strong correlation both between inner and outer barrel hit multiplicities and with the MBD, validating synchronization across all eight FELIX servers, and successful demonstration of streaming-readout capability. Timing studies show the analog timing resolution is $\sim13$ ns, well within the single beam-crossing window, and a low intrinsic noise level of $\sim0.02$ noise hits per bunch crossing.

Overall, the INTT has been successfully integrated into the sPHENIX data acquisition system and achieves stable, high-efficiency operation, fulfilling its role of providing precise single-event timing and contributing to charged-particle tracking and vertexing for sPHENIX physics measurements.
\label{Summary}

\section{Acknowledgments}
The authors express their gratitude to the members of the sPHENIX collaboration and the staff of the Physics and Instrumentation Divisions at BNL for their continuous support for the assembly, installation, and operation of the INTT barrel system. The authors also thank the Collider-Accelerator Department for surveying the geometry and installation of the INTT barrel.

This work was supported by the Office of Nuclear Physics in the Office of Science of the Department of Energy and the National Science Foundation (U.S.A.); the Ministry of Education, Culture, Sports, Science, and Technology (MEXT) and the Japan Society for the Promotion of Science (JSPS), including JSPS KAKENHI Grant Number JP26KJ1384 (Japan); the RIKEN Student Researcher Program for supporting students (Japan); the National Science and Technology Council and the Ministry of Education (Taiwan); and the National Research Foundation of Korea (NRF) grant funded by the Korea government (MSIT) (No. 2018R1A5A1025563) (Republic of Korea).
\label{Acknowledgments}

%%%%%%%%%%%%%%%%%%%%%%%%%%%%%%%%%%%%%%%%%%%%%%%%%%
%%%%%%%%%%%%%%%% Reference %%%%%%%%%%%%%%%%%%%%%%%
%%%%%%%%%%%%%%%%%%%%%%%%%%%%%%%%%%%%%%%%%%%%%%%%%%
% \bibliographystyle{elsarticle-num}
%\bibliographystyle{unsrturl}
\bibliographystyle{cas-model2-names}
\bibliography{INTT_Barrel_NIM_main}

@misc{Aidala:2012nz,
    author = "Aidala, C. and others",
    title = "{sPHENIX: An Upgrade Concept from the PHENIX Collaboration}",
    eprint = "1207.6378",
    archivePrefix = "arXiv",
    primaryClass = "nucl-ex",
    month = "7",
    year = "2012",
    url={https://arxiv.org/abs/1207.6378},
}

@ARTICLE{Hahn:2003nim,
    author = "Hahn, H. and others",
    title = "{The RHIC design overview}",   
    doi = "10.1016/S0168-9002(02)01938-1",
    journal = "Nucl. Instrum. Meth. A",
    volume = "499",
    pages = "245-263",
    year = "2003"
}

@ARTICLE{Adcox:2003nim,
    author = "Adcox, K. and others",
    title = "{PHENIX detector overview}",
    archivePrefix = "arXiv",
    primaryClass = "physics.ins-det",
    doi = "10.1016/S0168-9002(02)01950-2",
    journal = "Nucl. Instrum. Meth. A",
    volume = "499",
    pages = "469-479",
    year = "2014"
}

@article{Belmont:2023fau,
    author = "Belmont, Ron and others",
    title = "{Predictions for the sPHENIX physics program}",
    doi = "10.1016/j.nuclphysa.2024.122821",
    journal = "Nucl. Phys. A",
    volume = "1043",
    pages = "122821",
    year = "2024"
}

@article{Nakagawa2025,
SerialNo={11},
author={Akiba, Y. and others},
year={2026},
title={The ladder and readout cables of intermediate silicon strip detector for sPHENIX},
journal={Nucl. Instrum. Meth. A},
volume={1082},
 pages={171020},
post={\myehost{http://arxiv.org/abs/2503.09105}[arXiv:2503.09105]},
primaryClass={physics.ins-det},
doi={10.1016/j.nima.2025.171020},
}

@article{SHIH2026171312,
title = {Beam test results of the Intermediate Silicon Tracker for sPHENIX},
journal = {Nucl. Instrum. Meth. A},
volume = {1086},
pages = {171312},
year = {2026},
issn = {0168-9002},
doi = {https://doi.org/10.1016/j.nima.2026.171312},
url = {https://www.sciencedirect.com/science/article/pii/S0168900226000380},
author = {Shih, C.W. and others}
}

@ARTICLE{Aune:2024nim,
    author = "Aune, S. and others",
    title = "{The sPHENIX Micromegas Outer Tracker}",
    archivePrefix = "arXiv",
    primaryClass = "physics.ins-det",
    doi = "10.1016/j.nima.2024.169615",
    journal = "Nucl. Instrum. Meth. A",
    volume = "1066",
    pages = "169615",
    year = "2024"
}

@ARTICLE{Aubert2002nim,
    author = "Aubert, B. and others",
    title = "{The BABAR detector}",
    archivePrefix = "arXiv",
    primaryClass = "physics.ins-det",
    doi = "10.1016/S0168-9002(01)02012-5",
    journal = "Nucl. Instrum. Meth. A",
    volume = "479",
    pages = "1",
    year = "2002"
}

@article{Aidala:2013vna,
    author = "Aidala, C. and others",
    title = "{The PHENIX Forward Silicon Vertex Detector}",
    eprint = "1311.3594",
    archivePrefix = "arXiv",
    primaryClass = "physics.ins-det",
    doi = "10.1016/j.nima.2014.04.017",
    journal = "Nucl. Instrum. Meth. A",
    volume = "755",
    pages = "44--61",
    year = "2014"
}

@manual{SiliconHPK,
    author  = "{{Hamamatsu-Photonics K. K.}}",
    title = "model S14629-01 data sheet",
}

@INPROCEEDINGS{PHENIX_FPHX_cite,
  author={Hoff, J. R. and Zimmerman, T. N. and Yarema, R. J. and Kapustinsky, J. S. and Brookes, M. L.},
  booktitle={2009 IEEE Nuclear Science Symposium Conference Record (NSS/MIC)}, 
  title="{FPHX: A new silicon strip readout chip for the PHENIX experiment at RHIC}", 
  year={2009},
  volume={},
  number={},
  pages={75-79},
  doi={10.1109/NSSMIC.2009.5401873}}

@ARTICLE{Kapustinsky:2010nim,
    author = "Kapustinsky, J. S. and others",
    title = "{Production and performance of the silicon sensor and custom readout electronics for the PHENIX FVTX tracker}",
    archivePrefix = "arXiv",
    primaryClass = "physics.ins-det",
    doi = "doi.org/10.1016/j.nima.2009.09.036",
    journal = "Nucl. Instrum. Meth. A",
    volume = "617",
    pages = "546-548",
    year = "2010"
}

@manual{FPHX_Manual,
    author = "Zimmerman, T. and Hoff, J.",
    organization = "Fermilab",
    title = "FPHX Chip Documentation",
    year = {2009}
}

@manual{RCDAQ,
    author  = "{{Martin L. Purschke}}",
    title = "{RCDAQ, a lightweight yet powerful data acquisition system}",
    url = "https://www.phenix.bnl.gov/~purschke/rcdaq/rcdaq_doc.pdf"
}

@ARTICLE{BEX,
  title="{Development of Long and High-Density Flexible Printed Circuits}",
  author={Takashi Kondo and Kohei Fujiwara and Takashi Hachiya and Hikaru Imai and Miu Morita and Itaru Nakagawa and Naoyuki Sato and Masato Tsuruta and Daisuke Yanagawa},
  journal={Transactions of The Japan Institute of Electronics Packaging},
  volume={15},
  number={ },
  pages={E21-007-1-E21-007-10},
  year={2022},
  doi={10.5104/jiepeng.15.E21-007-1}
}

@ARTICLE{Felix,
    author = {A.~Paramonov},
    title = "{FELIX: the Detector Interface for the ATLAS Experiment at CERN}",
    journal = {EPJ Web of Conferences}, 
    volume  = {251},
    pages   = {04006},
    year = {2021},
    doi={10.1051/epjconf/202125104006}
}

@article{Anderson_2016,
doi = {10.1088/1748-0221/11/12/C12023},
url = {https://dx.doi.org/10.1088/1748-0221/11/12/C12023},
year = {2016},
month = {dec},
publisher = {},
volume = {11},
number = {12},
pages = {C12023},
author = {J. Anderson et al.},
title = "{FELIX: a PCIe based high-throughput approach for interfacing front-end and trigger electronics in the ATLAS Upgrade framework}",
journal = {Journal of Instrumentation}
}

@misc{LRP_2015_cite,
  author       = {USDOE Office of Science (SC) (United States)},
  title        = "{Reaching for the Horizon: The 2015 Long Range Plan for Nuclear Science}",
  url          = {https://www.osti.gov/biblio/1296778},
  place        = {United States},
  year         = {2015},
  month        = {09}}

@article{Abdulhamid2025,
  author    = {{The sPHENIX collaboration}},
  title     = "{Measurement of charged hadron multiplicity in Au+Au collisions at $\sqrt{s_{NN}} = 200$ GeV with the sPHENIX detector}",
  journal   = {Journal of High Energy Physics},
  volume    = {2025},
  number    = {8},
  pages     = {75},
  year      = {2025},
  doi       = {10.1007/JHEP08(2025)075},
  url       = {https://doi.org/10.1007/JHEP08(2025)075},
  issn      = {1029-8479}
}

@misc{Ignition,
  author       = {{Inductive Automation}},
  title        = {Ignition VISION Module},
  version      = {8.1},
  year         = {2026},
  url          = {https://inductiveautomation.com/ignition/modules/vision},
}

@misc{Grafana,
  author       = {{Grafana Labs}},
  title        = {Grafana},
  version      = {10.4},
  year         = {2026},
  url          = {https://grafana.com},
}

\end{document}